\documentstyle[12pt,epsfig]{article}

\def\beq{\begin{equation}}
\def\eeq#1{\label{#1}\end{equation}}
\def\eeqn{\end{equation}}

\def\beqa{\begin{eqnarray}}
\def\eeqa#1{\label{#1}\end{eqnarray}}
\def\eeqan{\end{eqnarray}}

\let\bar=\overbar

\def\O{{\cal O}}

\def\Dslash{\not{\hbox{\kern-4pt $D$}}}
\def\dslash{\not{\hbox{\kern-2pt $\del$}}}

\def\msb{{\bar{\ssstyle M \kern -1pt S}}}

\def\Title#1{\begin{center} {\Large {\bf #1} } \end{center}}

\begin{document}

\Title{Tests of Perturbative QCD and Jet Physics}

\bigskip\bigskip


\begin{raggedright}  

{\it J. Womersley\index{Womersley, J.}\\
Fermi National Accelerator Laboratory,\\
Batavia, Illinois 60510 }
\bigskip\bigskip
\end{raggedright}

Quantum Chromodynamics (QCD) is a vast domain. In the past twelve months,
no less than
473 papers with ``QCD'' in the title were submitted to the {\tt hep-ph}
archive (admittedly, this includes such titles as ``Cosmological QCD Phase
Transition and Dark Matter'').  More relevant, perhaps, is to note that
93 QCD-related abstracts were submitted to the 1999 Europhysics Conference
on High Energy Physics (EPS99). I shall therefore have to be selective,
and will organize this presentation by final state: jets, photons, weak 
bosons and heavy flavor.  

I would like to thank everyone who helped me put this presentation together,
and extend my apologies to all those whose work had to be omitted or
brutally summarized.  I also ask the audience's indulgence for any biases
from my particular background as a Tevatron experimenter. 

\section{Jets}
\index{jet}

\subsection{Inclusive Jet Cross Sections at $\sqrt{s}=1.8$~TeV}

\begin{figure}[tb]
\begin{center}
\includegraphics*[height=6cm]{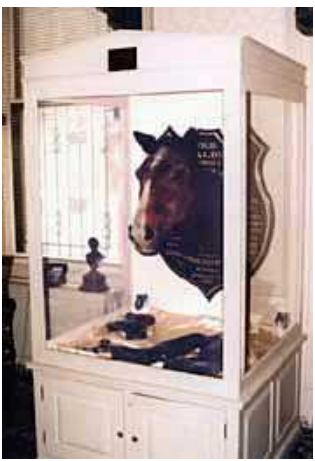}
\caption{``The horse is dead.''}
\label{fig:deadhorse}
\end{center}
\end{figure}

\begin{figure}[p]
\begin{center}
\begin{tabular}{cc}
\includegraphics*[bb=30 140 525 655,height=6cm]{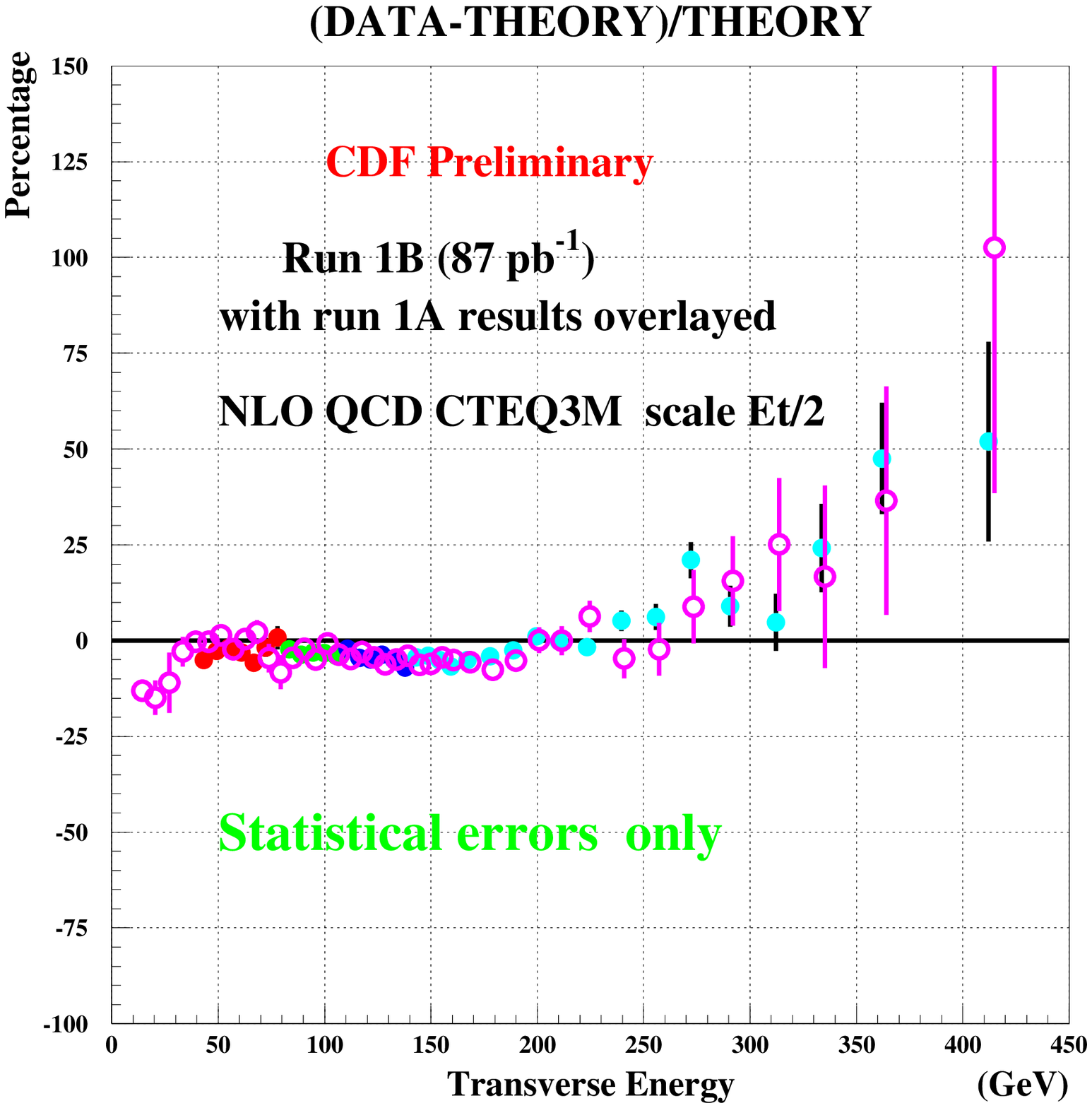}&
\includegraphics*[height=6cm]{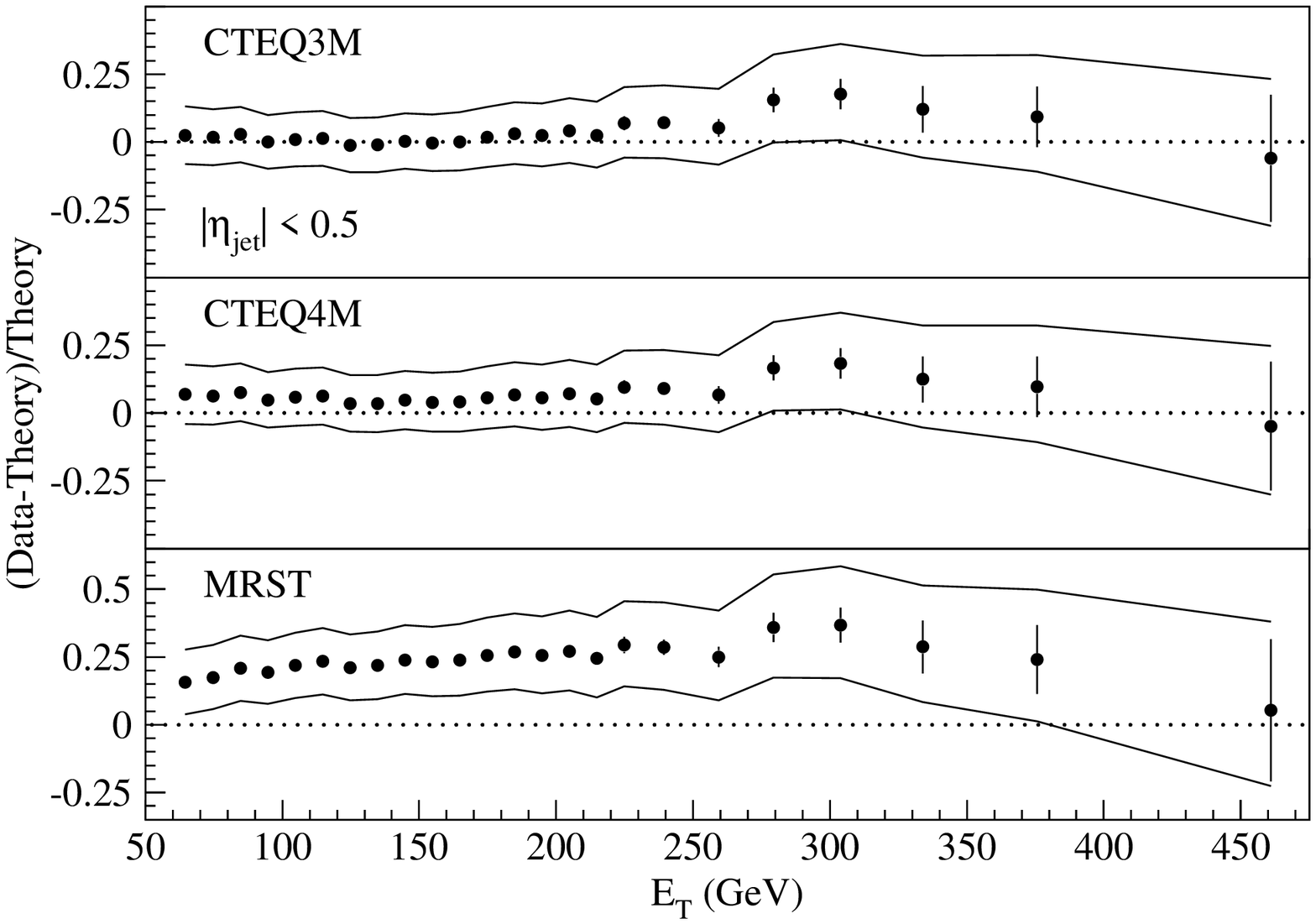}\\
\end{tabular}
\caption{Inclusive jet cross sections measured at the
Tevatron by CDF\protect\cite{cdfjets} 
(left, for $0.1<|\eta|<0.7$) and D\O\protect\cite{d0jets} 
(right, for $|\eta|< 0.5$), all normalized to the NLO QCD prediction.}
\label{fig:tevjets}

\vspace{2cm}
\includegraphics*[height=10cm]{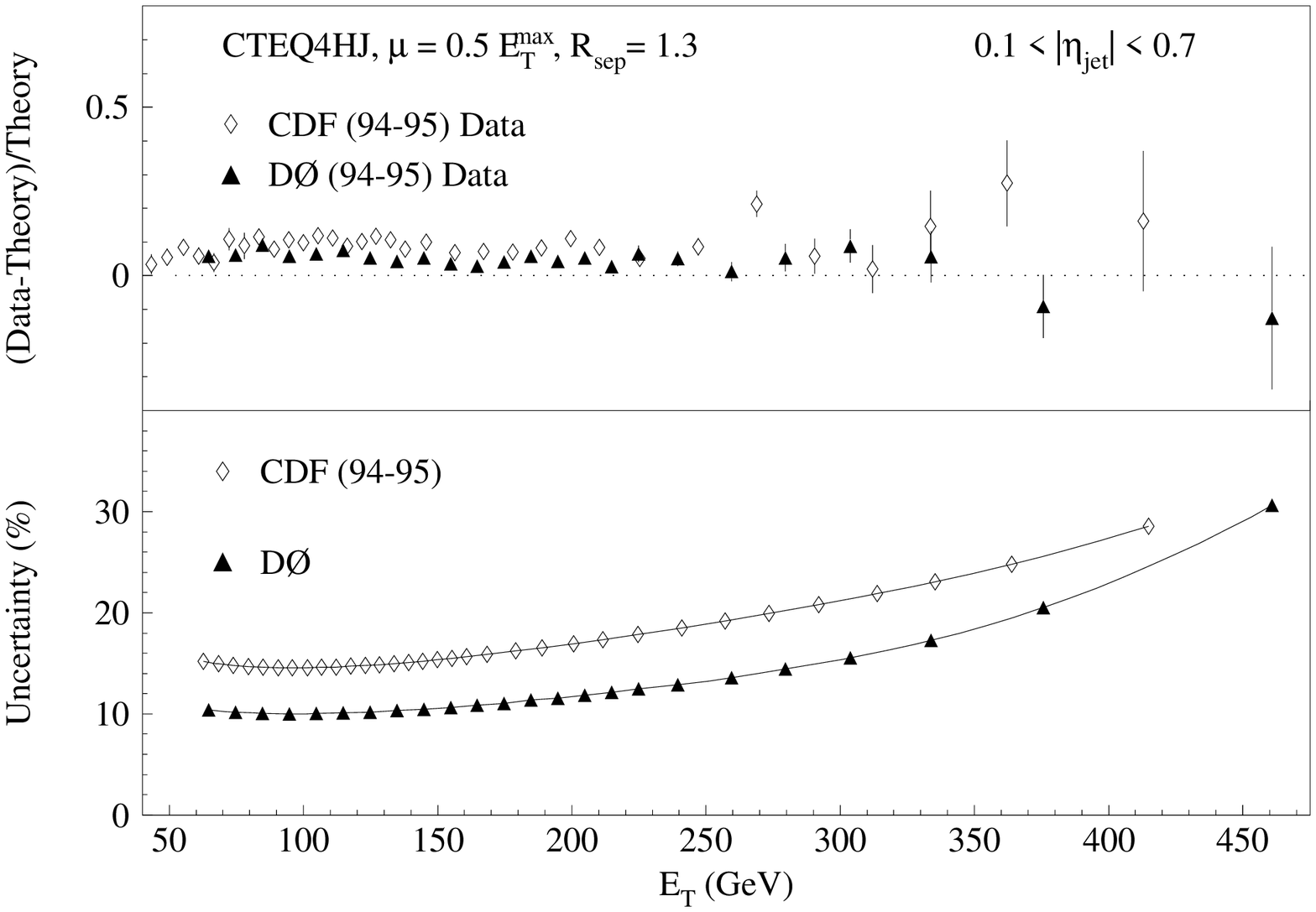}
\vspace{-2cm}
\caption{Inclusive jet cross sections for $0.1<|\eta|<0.7$
from CDF and D\O, compared with CTEQ4HJ distribution; and
size of the systematic errors on the two measurements.
Taken from \protect\cite{annrev}.} 
\label{fig:annrev}
\end{center}
\end{figure}

So much has been said about the high-$E_T$ behavior of the inclusive
jet cross section at the Tevatron that it is difficult to know what
can usefully be added (see Fig.~\ref{fig:deadhorse}).  
The measured central inclusive jet cross sections, from 
CDF\cite{cdfjets} and D{\O}\cite{d0jets},
compared with the NLO theory, are shown in Fig.~\ref{fig:tevjets}
(note that the CDF figure does not include systematic errors).
The impression one gets is that there is a marked excess above
QCD in the CDF data, which is not observed at D\O.


In order to compare with CDF, D\O\ carried out an analysis in exactly
the same rapidity interval ($0.1 < |\eta| < 0.7$).  The results\cite{annrev}
are shown in Fig.\ref{fig:annrev}.  Firstly we note that there is no 
actual discrepancy between the datasets. Secondly, for this plot the 
theoretical prediction was made using the CTEQ4HJ parton distribution,
which has been adjusted to give an increased gluon density at large $x$ while
not violating any experimental constraints (except perhaps fixed target photon
production data, which in any case
require big corrections before they can be compared to QCD, as we shall
see later).  The result of this increased gluon content is improved
agreement especially with the CDF data points.

What then have we learned from this issue?  In my opinion, whether the
CDF data show a real excess above QCD, or just a ``visual excess,''
depends critically on understanding the systematic errors and their
correlations as a function of $E_T$.  Whether nature has actually exploited
the freedom to enhance gluon distributions at large $x$ will only be clear 
with the addition of more data --- the factor of 20 increase in luminosity
in Run~II will extend the reach significantly in $E_T$ and should make
the asymptotic behavior clearer.  Whatever the Run~II data show, this has
been a useful lesson; it has reminded us all that parton distributions
have uncertainties, whether made explicit or not, and that a full 
understanding of experimental systematics {\it and their correlations} is needed
to understand whether experiments and theory agree or disagree.   

\begin{figure}[tb]
\begin{center}
\begin{tabular}{cc}
\includegraphics*[bb=30 140 525 655,height=6cm]{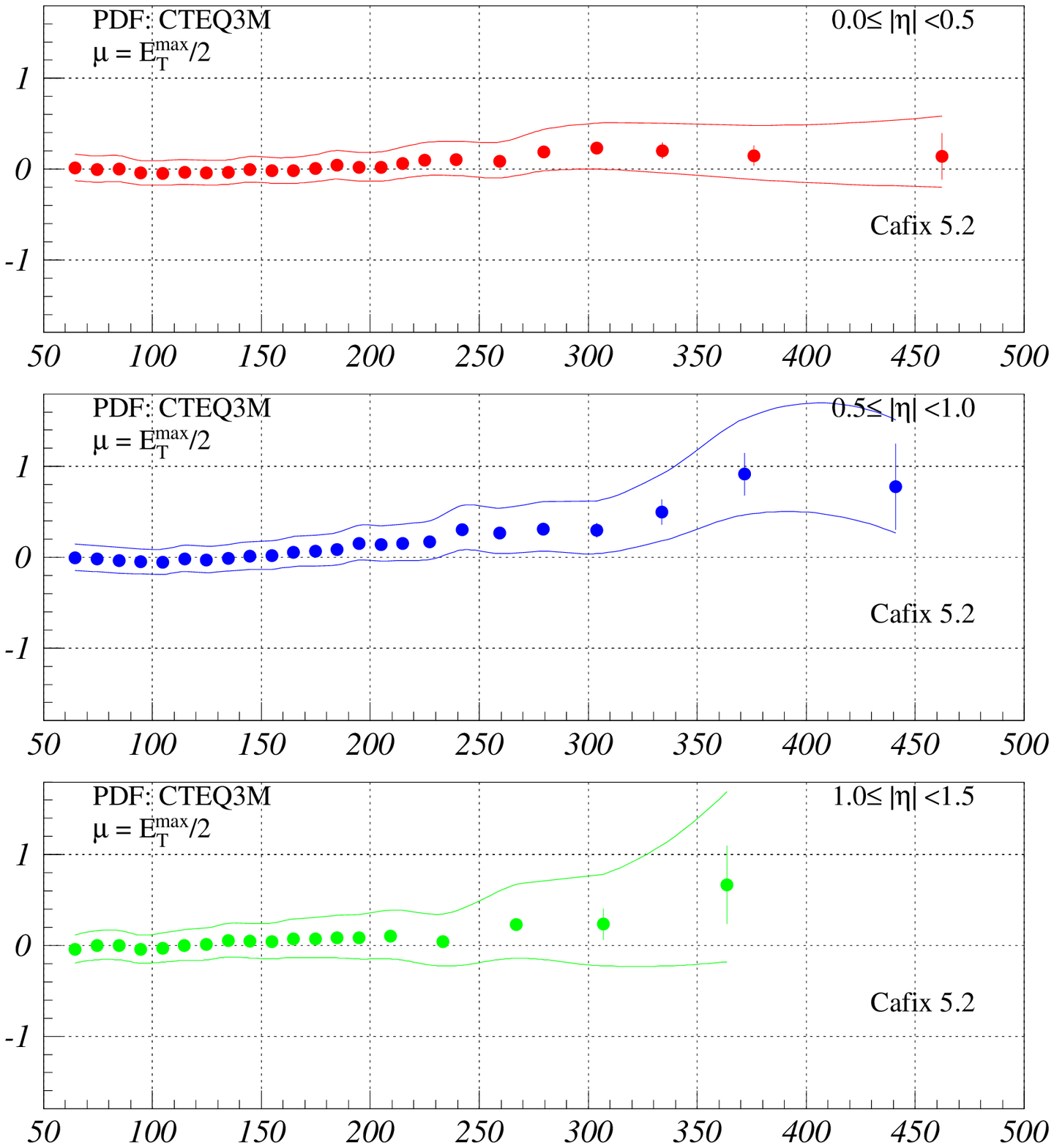}&
\includegraphics*[bb=30 140 525 655,height=6cm]{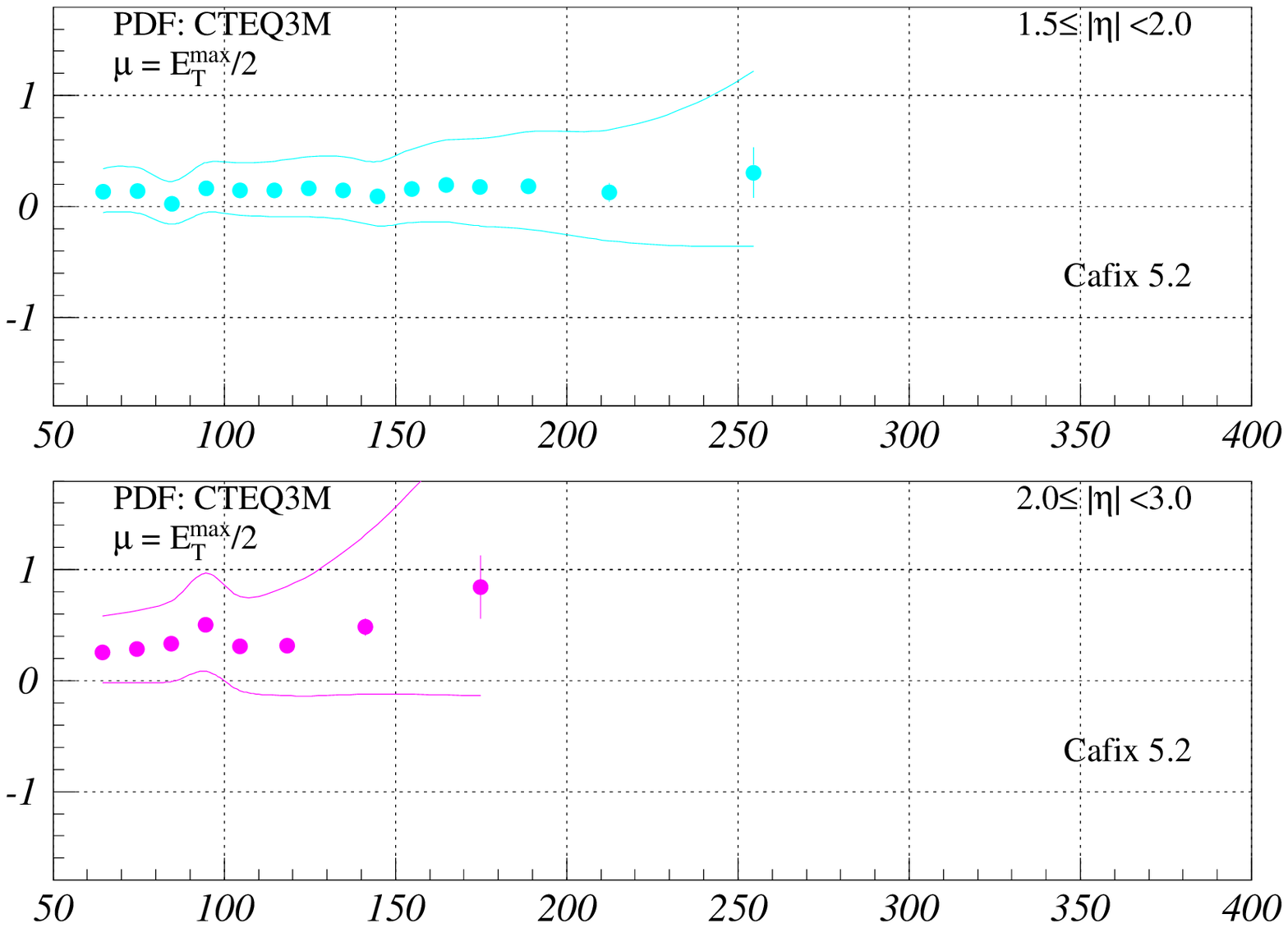}\\
\end{tabular}
\caption{Inclusive jet cross sections measured up to $|\eta|=3$ 
by D\O\protect\cite{d0fwdjets},
compared to the NLO QCD prediction (from JETRAD using the 
CTEQ3M parton distributions).
\label{fig:fwdjets}}
\end{center}
\end{figure}

D\O\cite{d0fwdjets} 
have extended their measurement of inclusive jet cross sections into
the forward region.  Figure~\ref{fig:fwdjets} shows the measured cross
sections up to  $|\eta|=3$.  They are in good agreement with NLO
QCD over the whole range of pseudorapidity and transverse energy.

\begin{figure}[tb]
\begin{center}
\begin{tabular}{cc}
\includegraphics*[height=6.5cm]{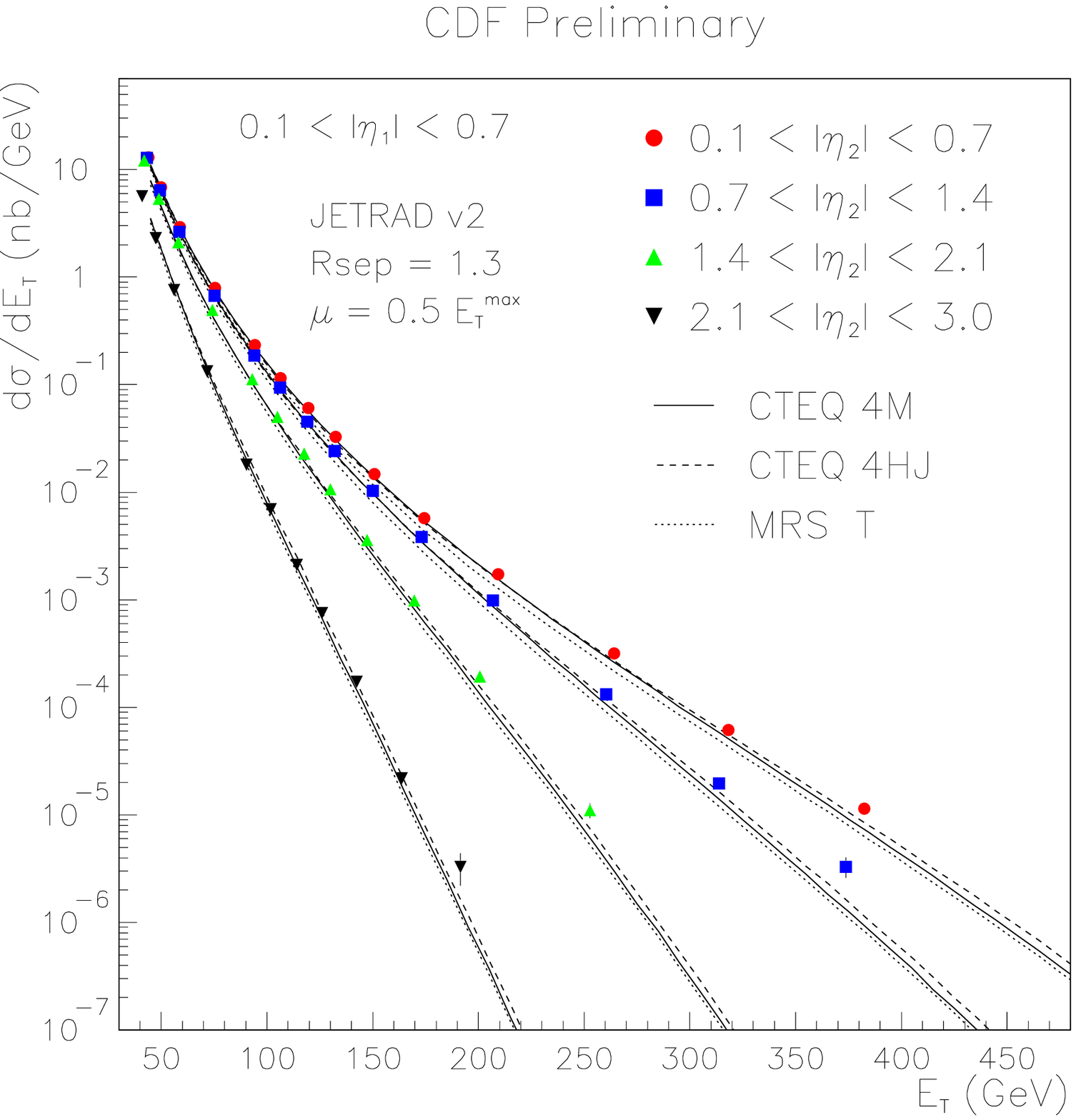}&
\includegraphics*[height=6.5cm]{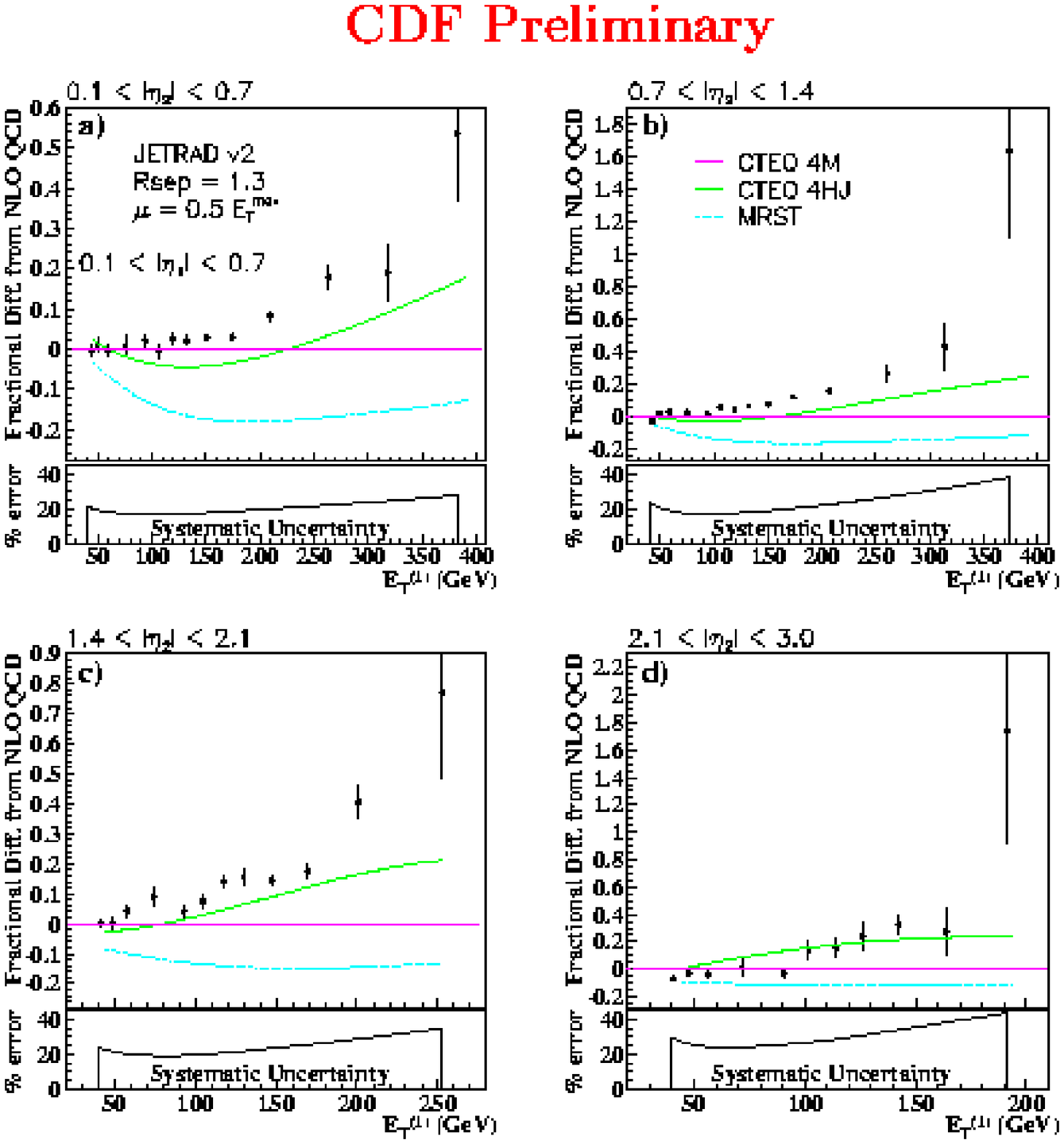}\\
\end{tabular}
\caption{Dijet cross sections measured
by CDF\protect\cite{cdftripdiff} for events with one central
jet $0.1<|\eta_1|<0.7$ and one jet allowed forward;
left, as a function of the central jet $E_T$ for various bins
of $|\eta_2|$, and right, normalized to the NLO QCD prediction
(from JETRAD).}
\label{fig:cdftripdiff}
\end{center}
\end{figure}

\subsection{Dijet Production}

Both Tevatron experiments have also studied dijet final states.
CDF\cite{cdftripdiff} has presented cross sections for processes
with one central jet ($0.1 < |\eta_1| < 0.7$) and one jet allowed
forward ($|\eta_2|$ up to 3.0).  In Fig.~\ref{fig:cdftripdiff} these
are compared with the NLO QCD prediction ast a function of the 
central jet's transverse energy ($E_{T1}$). 
The data show an excess above the theory for large $E_{T1}$,
just as seen in the inclusive cross section; but since these events
are common to both samples, this is not surprising.

\begin{figure}[tb]
\begin{center}
\begin{tabular}{cc}
\includegraphics*[height=6cm]{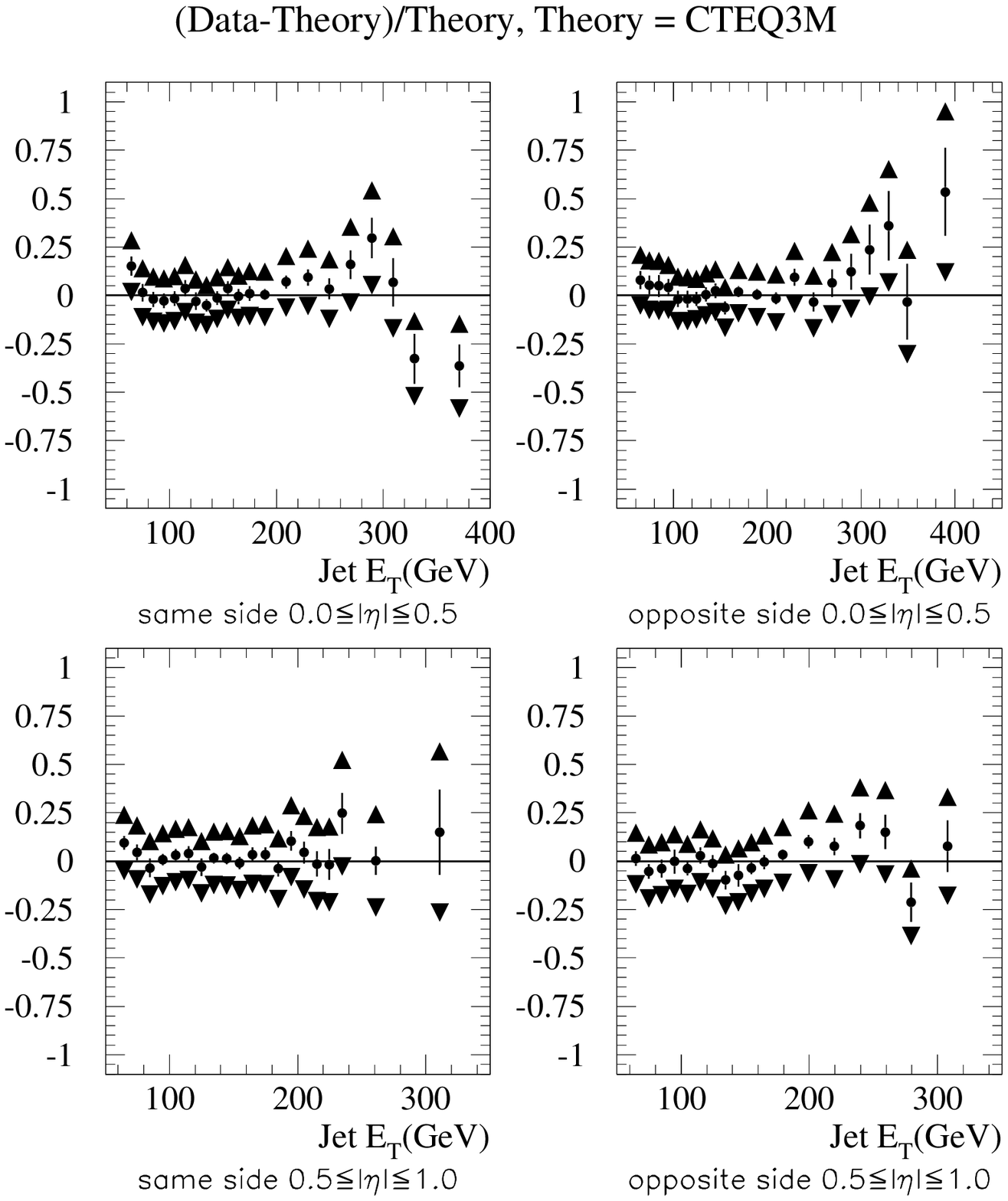}&
\includegraphics*[height=6cm]{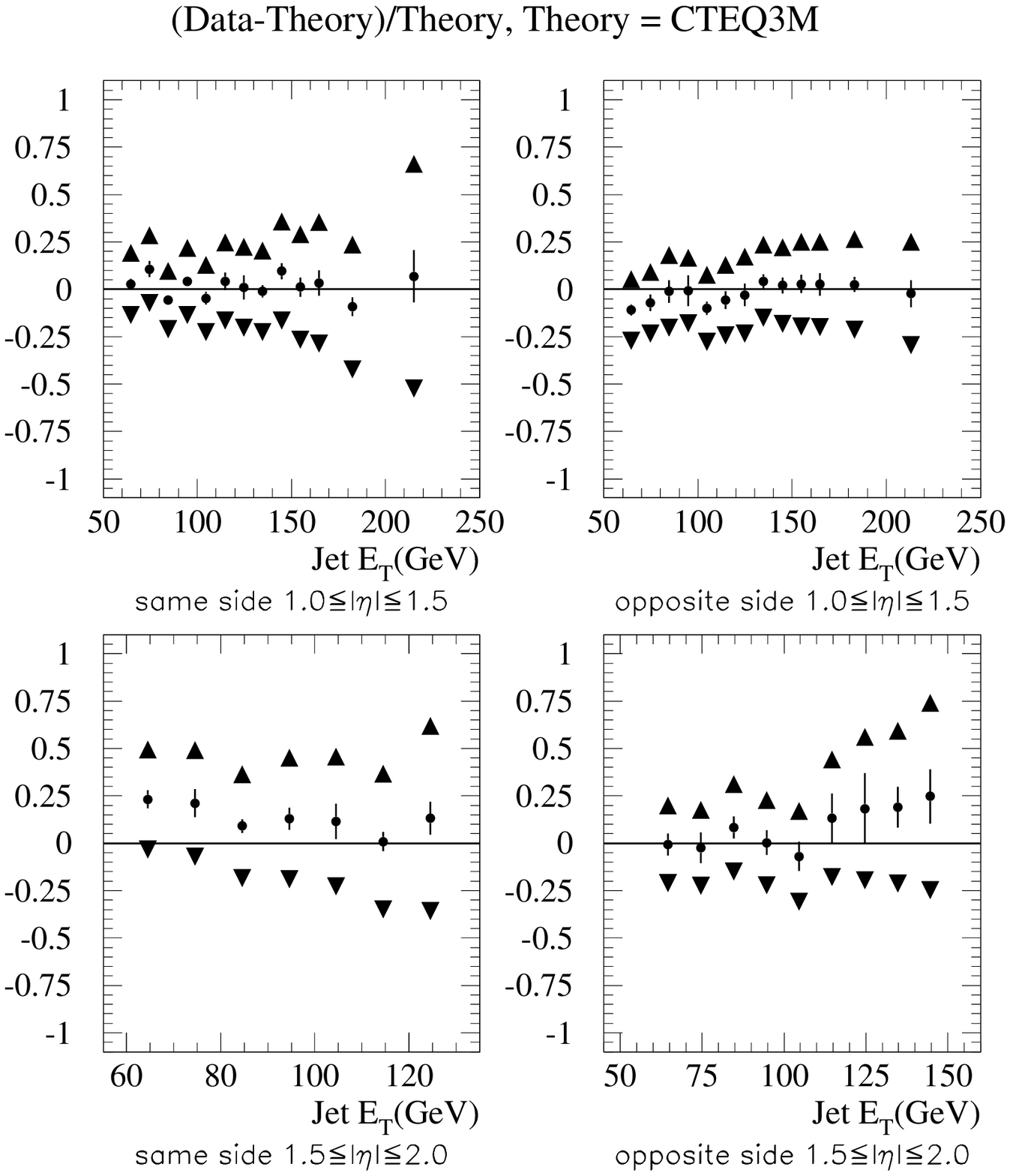}\\
\end{tabular}
\caption{Dijet jet cross sections measured up to $|\eta|=2$ 
by D\O\protect\cite{d0fwdjets}, for same side and opposite side
topologies,
compared to the NLO QCD prediction (from JETRAD using the 
CTEQ3M parton distributions).}
\label{fig:ssosjets}
\end{center}
\end{figure}

D\O\ have measured\cite{d0fwdjets} 
the cross sections for dijet production with both 
same-side ($\eta_1 \approx \eta_2$)
and opposite-side ($\eta_1 \approx - \eta_2$) topologies, for four
bins of $|\eta|$ up to 2.0.  The results are all in good agreement
with the NLO QCD prediction, as seen in Fig.~\ref{fig:ssosjets}.

\subsection{Jet Cross Sections at $\sqrt{s}=630$~GeV}

\begin{figure}[p]
\begin{center}
\includegraphics*[height=8cm]{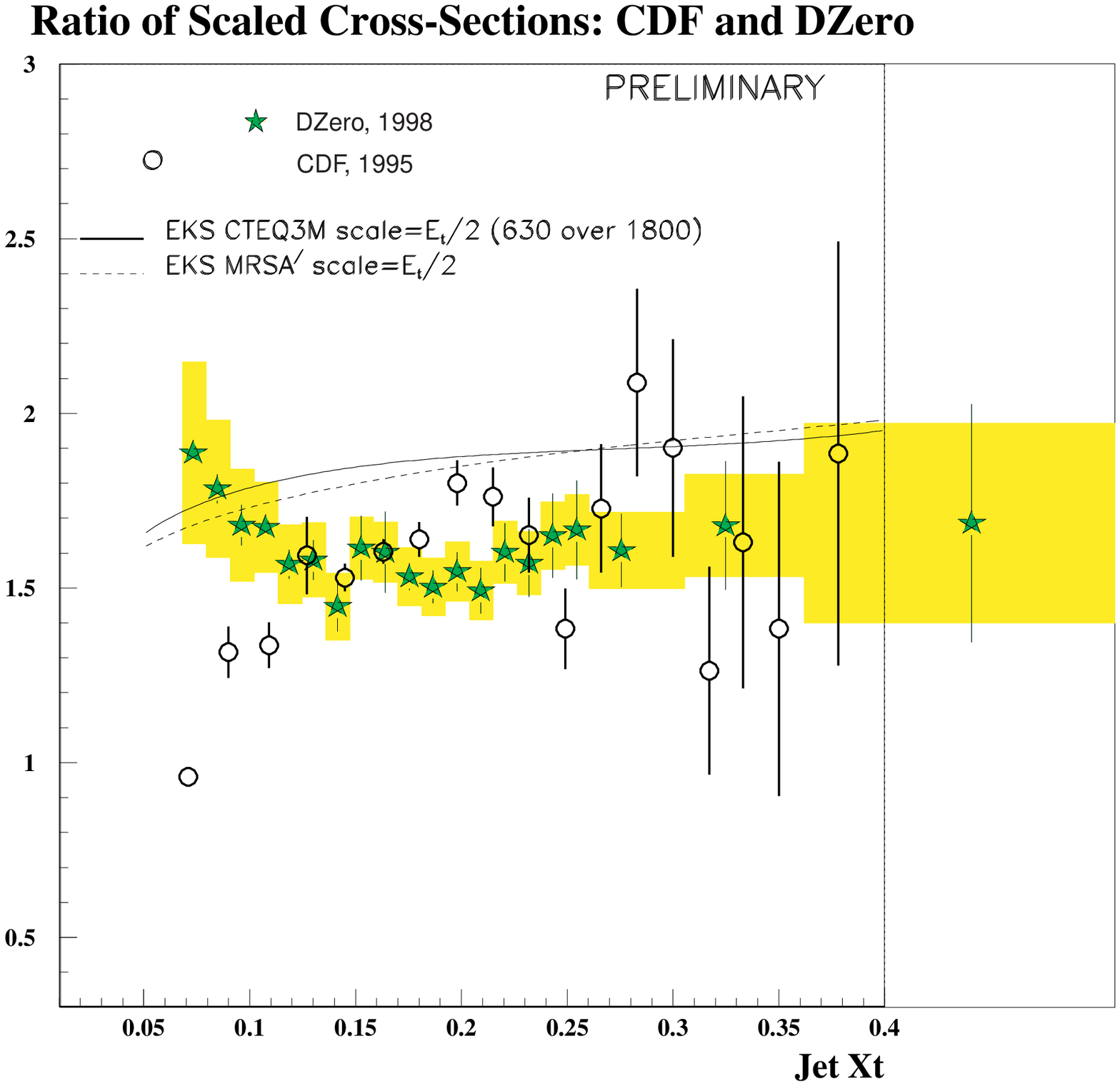}
\caption{Ratio of jet cross sections at $\sqrt{s}=1800~GeV$ 
to $\sqrt{s}=630~GeV$, as a function of $x_T = 2E_T/\sqrt{s}$,
as measured by CDF\protect\cite{cdf630} and D\O\protect\cite{d0630} 
and as predicted by NLO QCD}
\label{fig:630}

\includegraphics*[height=8cm]{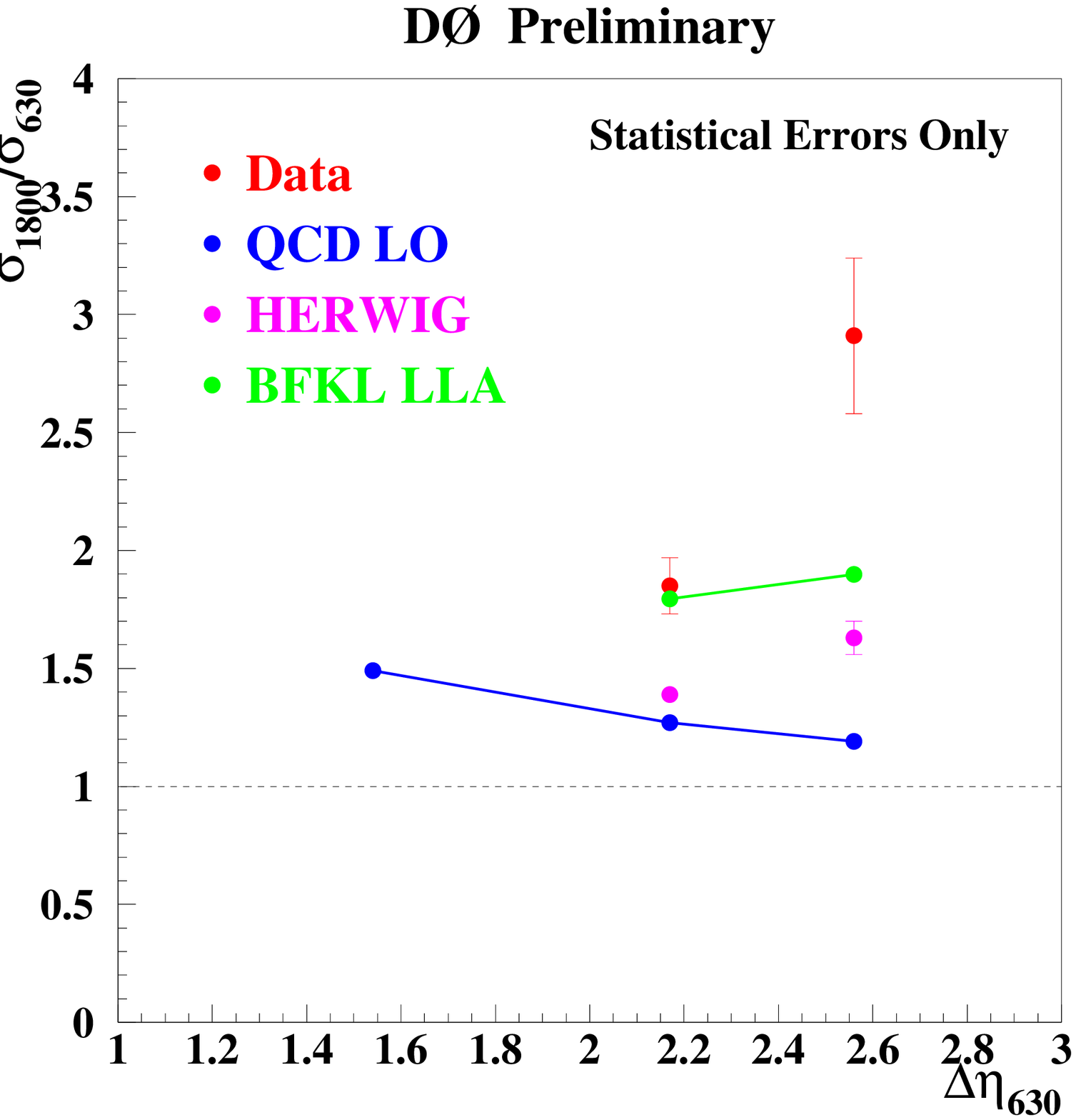}
\caption{Ratio of cross section at $\sqrt{s}=1800$~GeV to that at
$\sqrt{s}=630$~GeV, for events with dijets widely separated in 
rapidity\protect\cite{d0bfkl}.
\label{fig:bfkl}}
\end{center}
\end{figure}

Both CDF\cite{cdf630} and D\O\cite{d0630} have measured the
ratio of jet cross sections at $\sqrt{s}=1800$ and 630~GeV,
exploiting a short period of data taking at the latter center of mass energy
at the end of Run~I.  This ratio is expected to be a rather straightforward
quantity to measure and to calculate.  The ratio as a function
of scaled jet transverse energy $x_T = 2E_T/\sqrt{s}$ is shown in
Fig.\ref{fig:630}.  Unfortunately the two experiments are not obviously
consistent with each other (especially at low $x_T$) or with NLO QCD
(at any $x_T$).  At least two explanations have been suggested for
the discrepancy.
Firstly, different renormalization scales could be used for the
theoretical calculations at the two energies.  While allowed, this
seems unappealing. Glover has suggested that such
a procedure is in fact natural when a scaling variable
like $x_T$ is used; because $x_T$ differs by a factor of about 
three between the two center of mass energies for a given $E_T$,
a factor of three difference in the renormalization scales is appropriate. 
An alternative explanation is offered by Mangano, who notes that a shift
in jet energies by of order 3~GeV which might arise from non-perturbative
effects would bring the data in line with the prediction.  Such effects
might include losses outside the jet cone, underlying event energy, and 
intrinsic transverse momentum of the incoming partons; one might be
under or over-correcting the data and the two experiments might even obtain
different results depending on how the corrections were done (based on
data or Monte Carlo, for example).  It seems that more work, both
theoretical and exerimental, is needed before this question can be resolved.

\subsection{BFKL}

D\O\cite{d0bfkl} 
have used the 630~GeV data to make a measurement which a cynic might perhaps 
describe as ``yet another attempt to find an observable which displays
BFKL behaviour.''  The ratio of cross sections at 1800 and 630~GeV is
measured for dijets with large rapidity separations, using bins such
that $x$ and $Q^2$ are the same in the two datasets (and therefore the parton
distributions cancel, to first order).  Figure~\ref{fig:bfkl} shows this
ratio as a function of the rapidity separation $\Delta\eta$ at 630~GeV.  
The ratio is much
larger than unity, and rises with rapidity separation, qualitatively
like the BFKL expectation --- but also 
like HERWIG.  Maybe we are indeed seeing BFKL dynamics,
but given that we apparently can't predict the ratio of inclusive cross
sections between the two energies, I would caution against inferring 
too much.

\subsection{Jet Structure at the Tevatron}

\begin{figure}[tb]
\begin{center}
\includegraphics*[height=5cm]{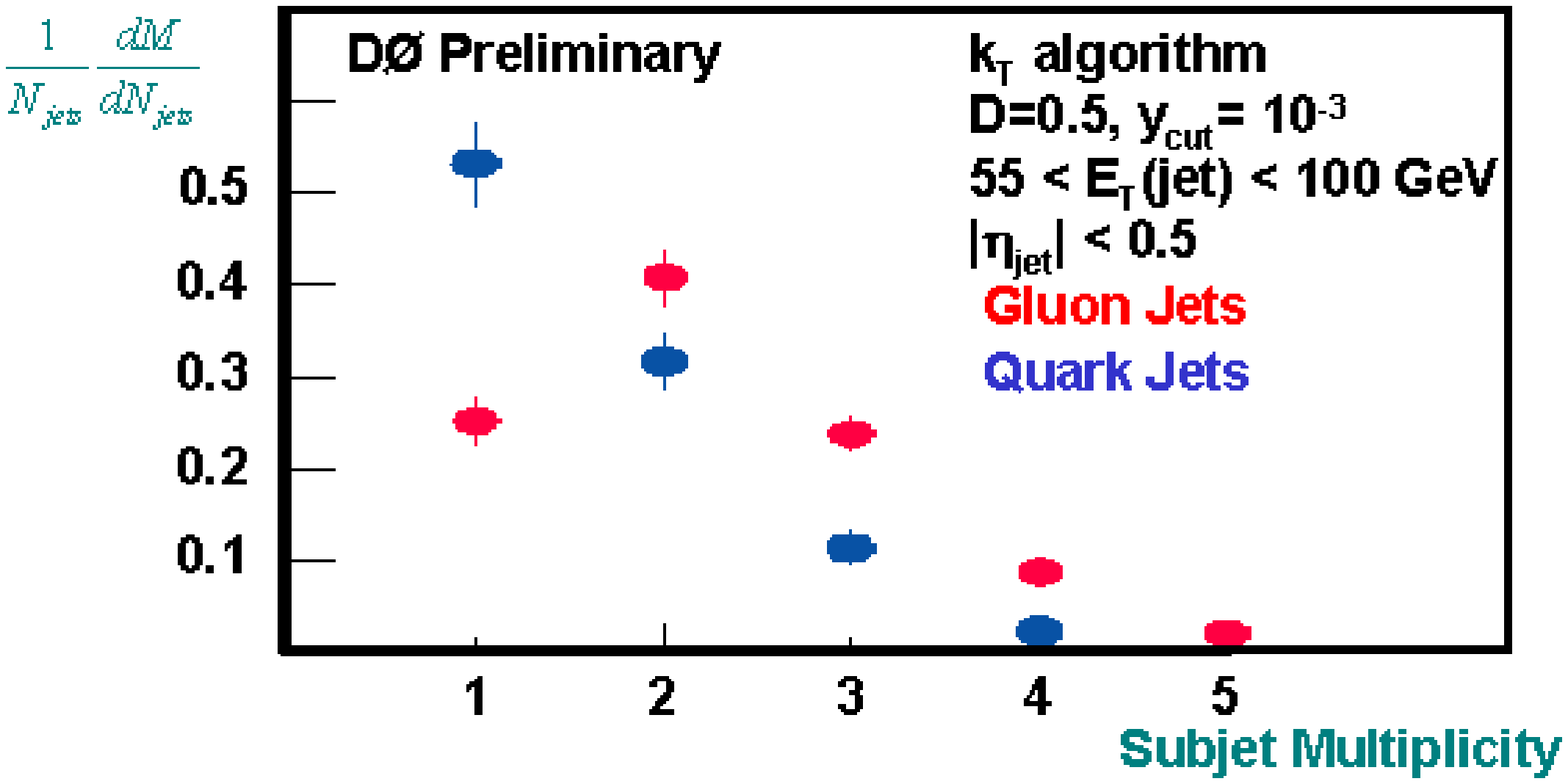}
\caption{Subjet multiplicities measured by D\O\ using a 
$k_T$ algorithm to find clusters within jets; distributions 
for quark and gluon
jets are inferred using $\sqrt{s}=1800$~GeV and
$\sqrt{s}=630$~GeV data\protect\cite{d0snihur}.
\label{fig:subjets}}
\end{center}
\end{figure}

All the results presented so far have used a cone jet finder.
By running a $k_T$ jet finder inside previously identified jets,
one can count the number of ``subjets'' or energy clusters.  This
allows the perturbative part of fragmentation to be studied.
D\O\cite{d0snihur} have made such a measurement and, by comparing
jets of the same $E_T$ and $\eta$ at $\sqrt{s}=1800$ and 630~GeV,
have inferred the composition of quark and gluon jets.
The extracted subjet multiplicity $M$ for the two species is shown
in Fig.\ref{fig:subjets}.  The ratio of $M-1$ for the two cases,
which might naively be expected to equal the ratio of gluon and
quark colour charges,
is found to be $1.91\pm0.04$, compared with $1.86\pm0.04$ from HERWIG.

These last three measurements all relied on a very short period of
Tevatron running at 630~GeV, which was surprisingly productive.  Though
it is politically difficult to get reduced energy running time when
one is searching for new physics, I suspect we may want to do
something like this once again in the next run. 

\subsection{Jet Production at HERA}

Many studies of jet production have also been carried 
out at HERA, by H1 and ZEUS\cite{herajets}.  
Inclusive jet, dijet and three-jet
cross sections have been measured as a function of jet $E_T$,
pseudorapidity and $Q^2$, typically using a $k_T$ jet finder.
The results are in good agreement with NLO (where available) or
LO QCD. ZEUS have also reported a study of subjets
similar to that described above\cite{herasubjets}; they find that the subjet 
multiplicity rises as the jet becomes more forward, which is
consistent with the expectation that more gluon jets would
be produced in this region of phase space (and also
consistent with the HERWIG Monte Carlo). 

\section{Direct Photon Production}
\index{photon}

\begin{figure}[p]
\begin{center}
\begin{tabular}{cc}
\includegraphics*[height=6.5cm]{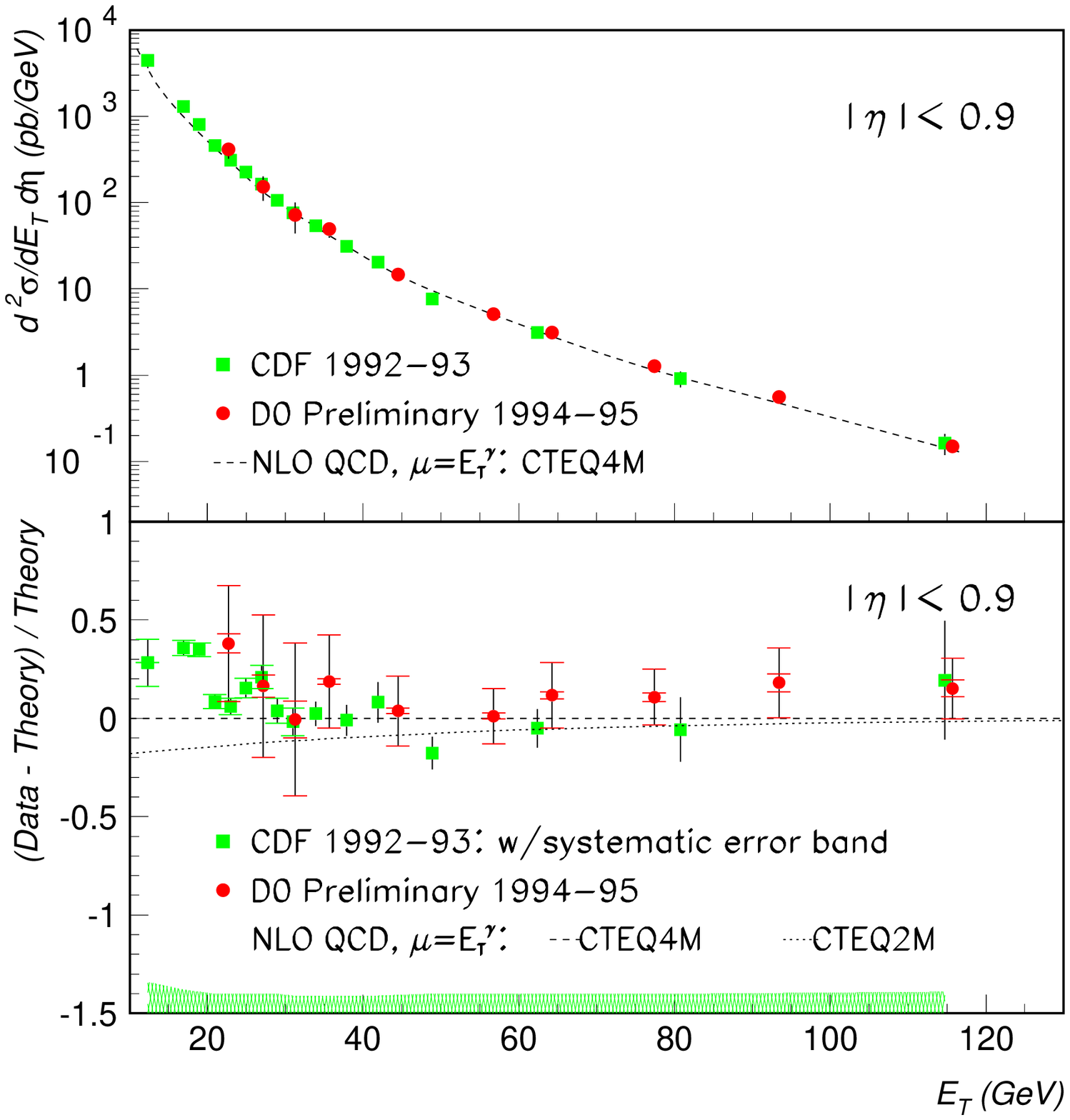}&
\includegraphics*[bb=30 140 540 655,height=6cm]{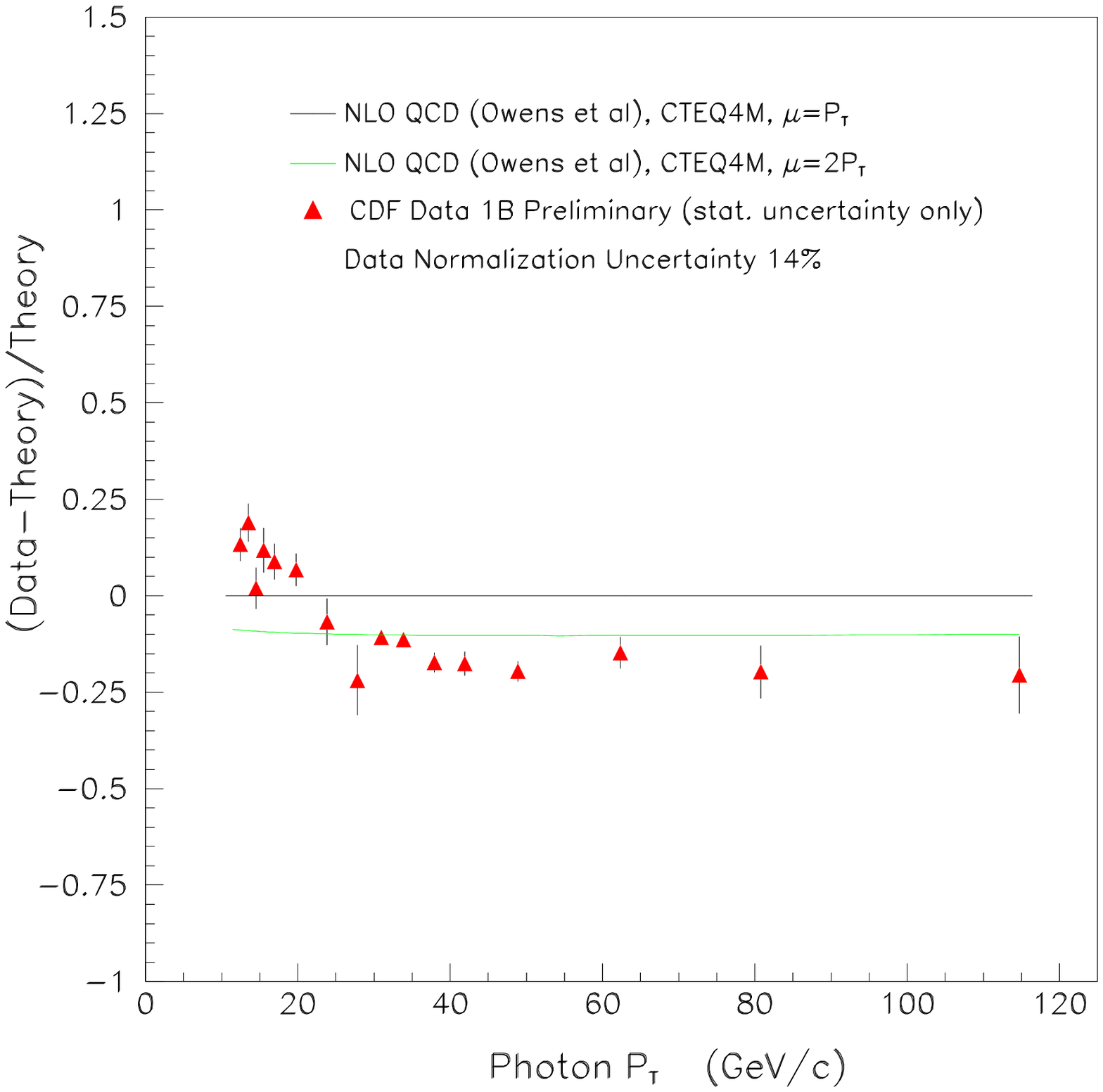}\\
\end{tabular}
\caption{Inclusive isolated direct photon cross sections at the Tevatron;
the left hand plot shows CDF\protect\cite{cdfphotons} 
and D\O\protect\cite{d0photons} measurements and the right hand plot
shows the latest CDF results\protect\cite{cdfnewphotons} 
(statistical errors only).  All are compared
with the NLO QCD prediction of Owens {\it et al}.\protect\cite{owens}.
\label{fig:photons}}
\vspace{1cm}
\begin{tabular}{cc}
\includegraphics*[bb=30 140 525 655,height=6.5cm]{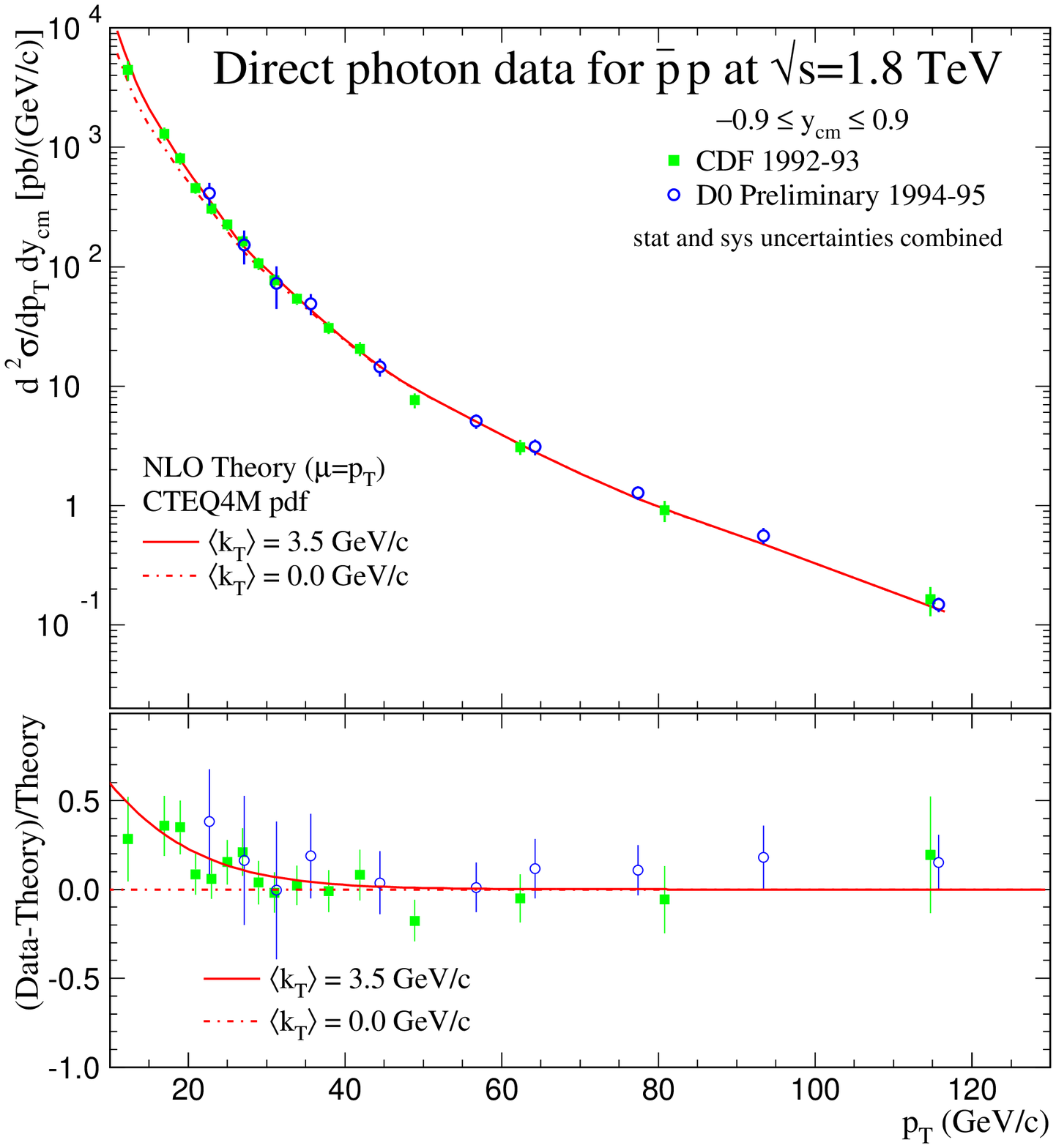}&
\includegraphics*[bb=30 140 525 655,height=6.5cm]{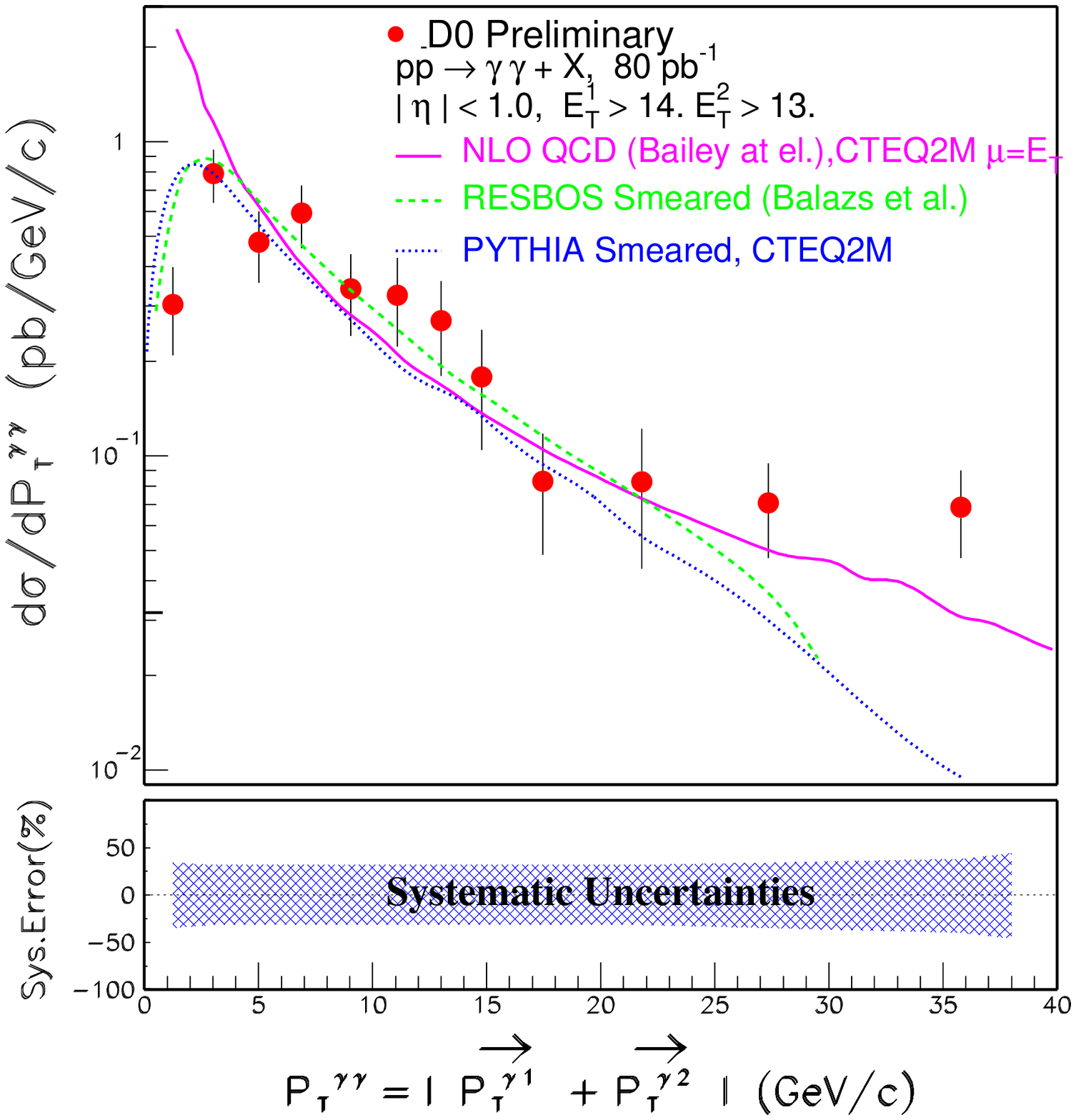}\\
\end{tabular}
\caption{The isolated photon cross section at the Tevatron (left hand plots)
agrees rather better
with QCD if 3.5~GeV of transverse momentum smearing 
(``$k_T$'') is added to account for soft gluon emission.  The magnitude of
the $k_T$ smearing is consistent with the most probable momentum of
$\gamma\gamma$ pairs (right hand plot)\protect\cite{diphotons}.
\label{fig:kttev}}
\end{center}
\end{figure}

Historically, many authors hoped that measurements of direct (or prompt) 
photons would provide a clean test of QCD, free from the 
systematic errors associated
with jets, and would help pin down parton distributions.  In fact photons
have not lived up to this promise --- instead they revealed that 
there may be unaccounted-for effects in QCD cross sections at 
low $E_T$.  (Because photons can typically be measured 
at lower energies than jets, they provide a way of exploring the
low-$E_T$ regime).  Results from the Tevatron experiments
\cite{cdfphotons}\cite{d0photons}\cite{cdfnewphotons} are shown in 
Fig.~\ref{fig:photons}. While the general agreement with the NLO calculation
of Owens and collaborators\cite{owens} is good,
there is a definite tendency for the data to rise above the theory
at low energies.  

An often-invoked explanation for this effect is that there is extra
transverse momentum smearing of the partonic system due to soft gluon
radiation.  The magnitude of the smearing, or ``$k_T$'', is typically
a few GeV (at the Tevatron), motivated in part by the experimentally
measured $p_T$ of the $\gamma\gamma$ system in diphoton production 
which peaks around 3~GeV\cite{diphotons}.  
Inclusion of such $k_T$ as a Gaussian
smearing in the calculation gives much better agreement with the data,
as shown in Fig.~\ref{fig:kttev}.

\begin{figure}[tb]
\begin{center}
\includegraphics*[bb=30 140 540 655,height=6.5cm]{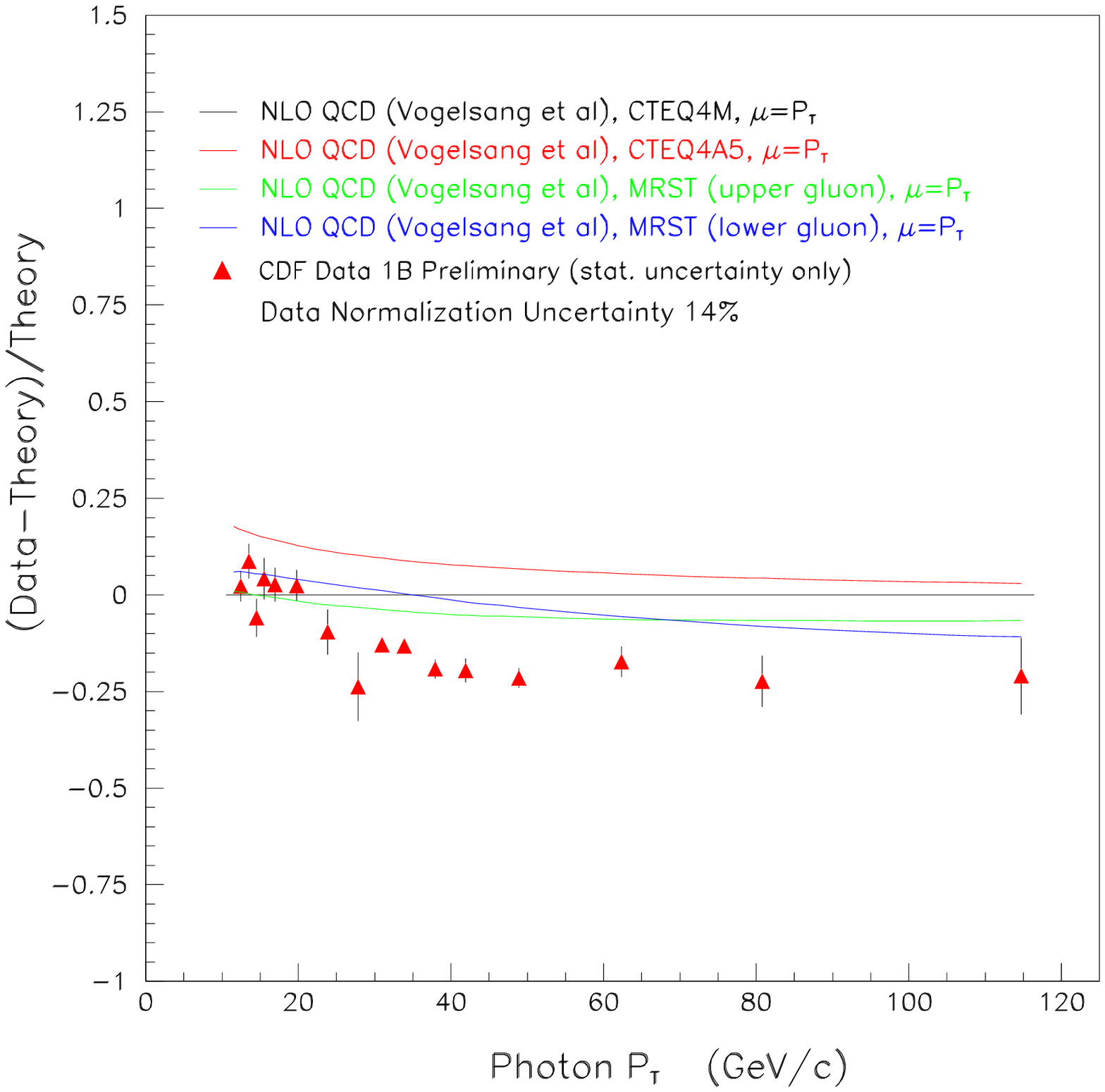}
\caption{Latest CDF isolated photon cross section normalized to the NLO
QCD prediction of Vogelsang {\it et al.}\protect\cite{vogelsang}, 
showing various choices of PDF and
renormalization scale within this prediction.  Statistical errors only.
\label{fig:vogelsang}}
\end{center}
\end{figure}

One should note that a shape a bit closer to that observed in
the data can be obtained in other ways. Vogelsang {\it et al.}\cite{vogelsang}
use a NLO treatment of fragmentation and allow the renormalization 
and factorization scales to differ, yielding the curves shown in
Fig.\ref{fig:vogelsang} without any $k_T$.  

\begin{figure}[p]
\begin{center}
\begin{tabular}{cc}
\includegraphics*[bb=30 140 525 655,height=6.5cm]
{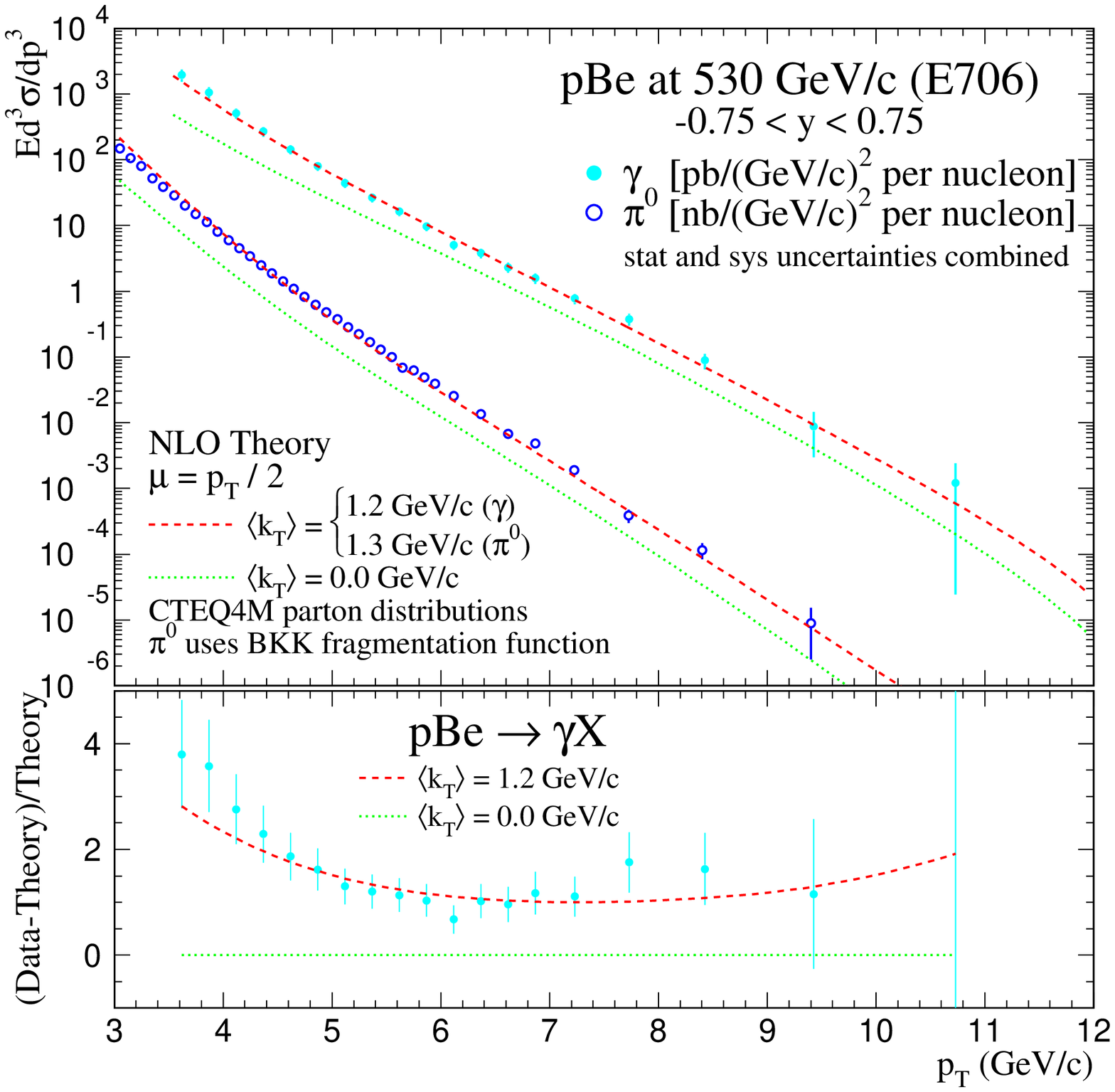}&
\includegraphics*[bb=30 140 525 655,height=6.5cm]{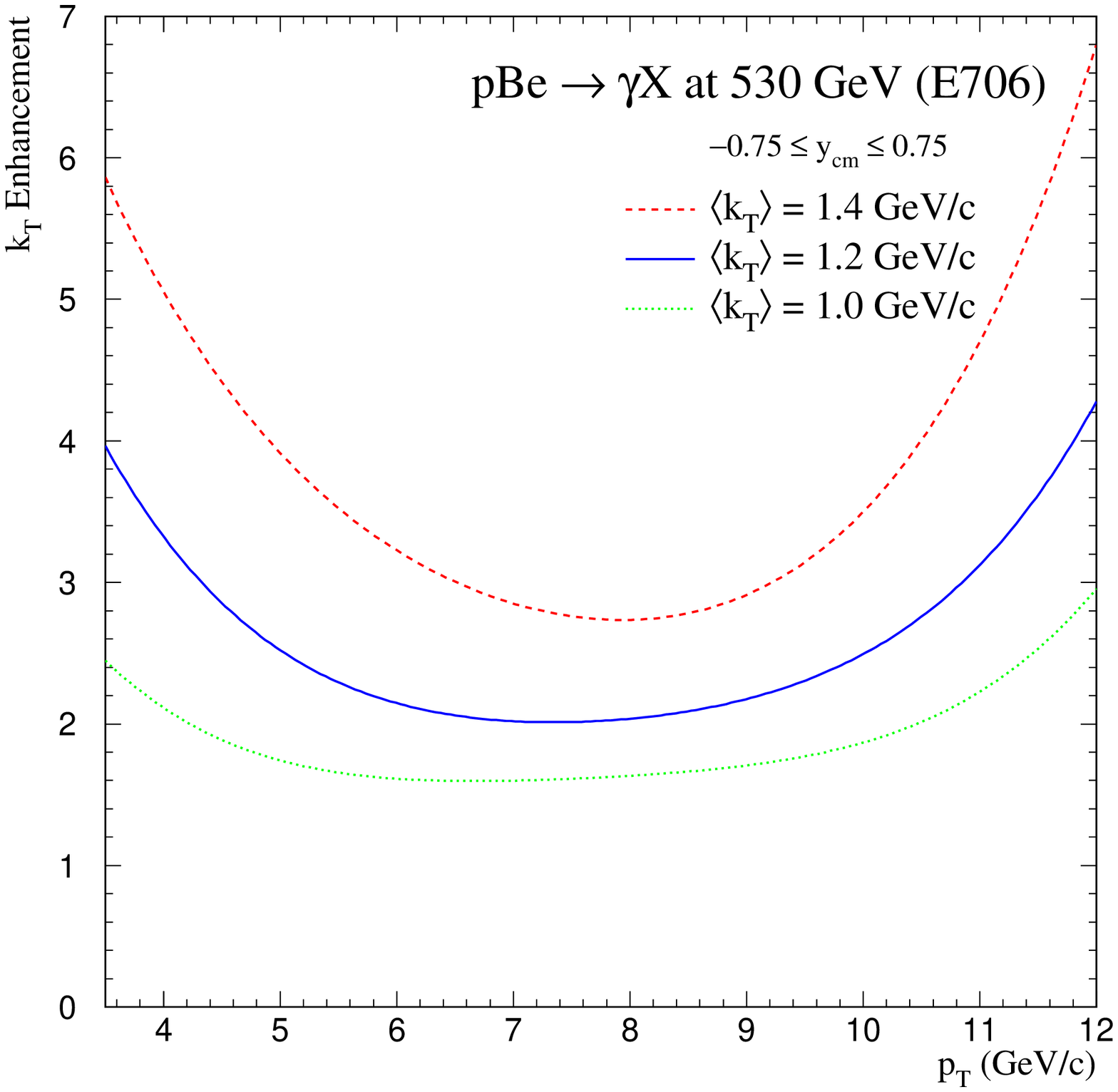}\\
\end{tabular}
\caption{The left hand plot shows the isolated photon and $\pi^0$ cross 
sections measured by E706\protect\cite{e706}, 
compared with the NLO QCD prediction with and 
without additional $k_T$ smearing.  The right hand plot shows the increase
in the cross section obtained when various values of $k_T$ are used.  
\label{fig:kt706}}

\includegraphics*[height=8cm]{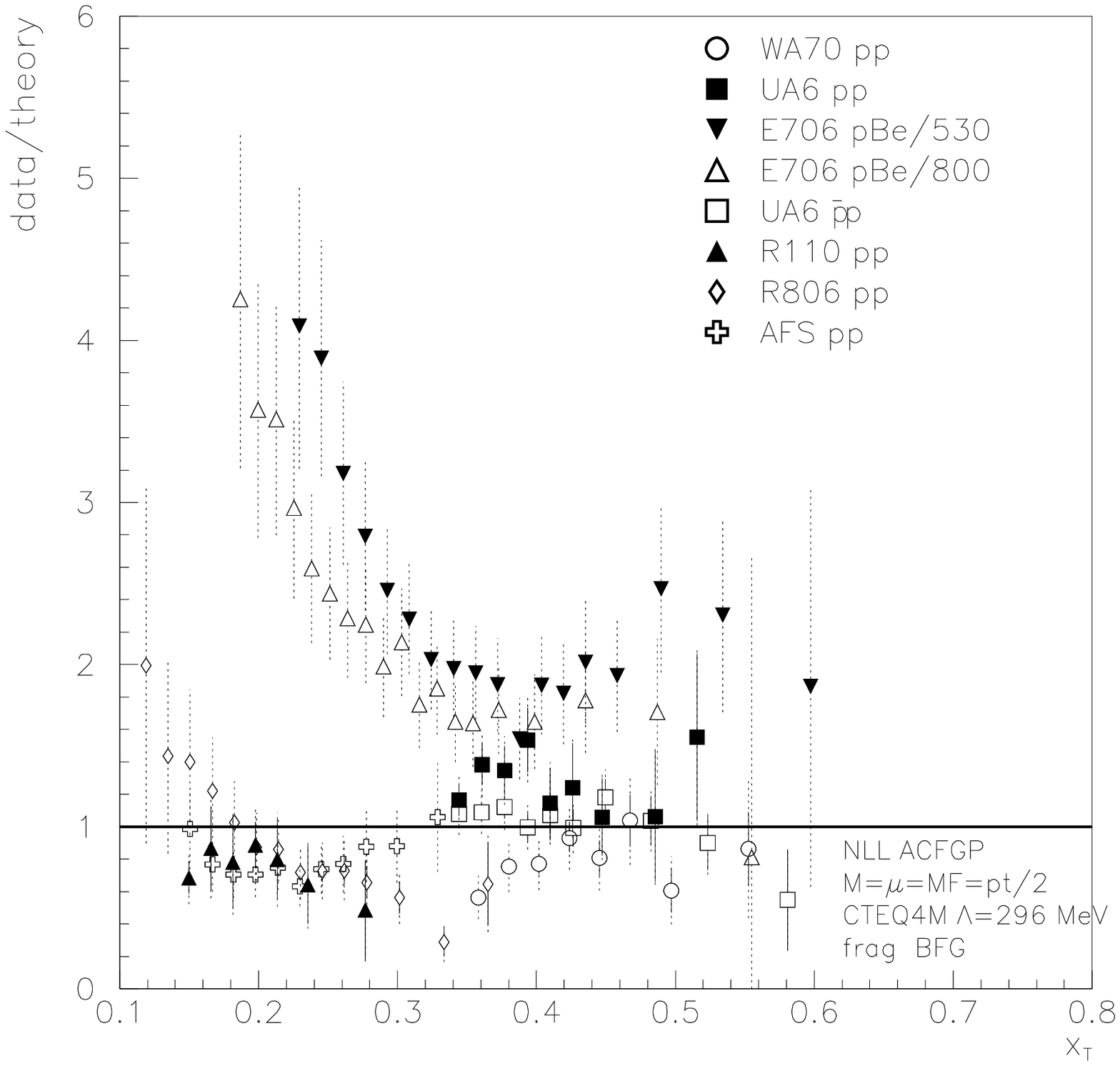}
\caption{Compilation of isolated photon cross sections compared with
NLO QCD, as a function of $x_T = 2 E_T/\sqrt{s}$.  From Aurenche {\it
et al.}\protect\cite{aurenche}
\label{fig:aurenche}}
\end{center}
\end{figure}

\begin{figure}[tb]
\begin{center}
\begin{tabular}{cc}
\includegraphics*[height=5.5cm]{e706_resum_comp.eps}&
\includegraphics*[height=6cm]{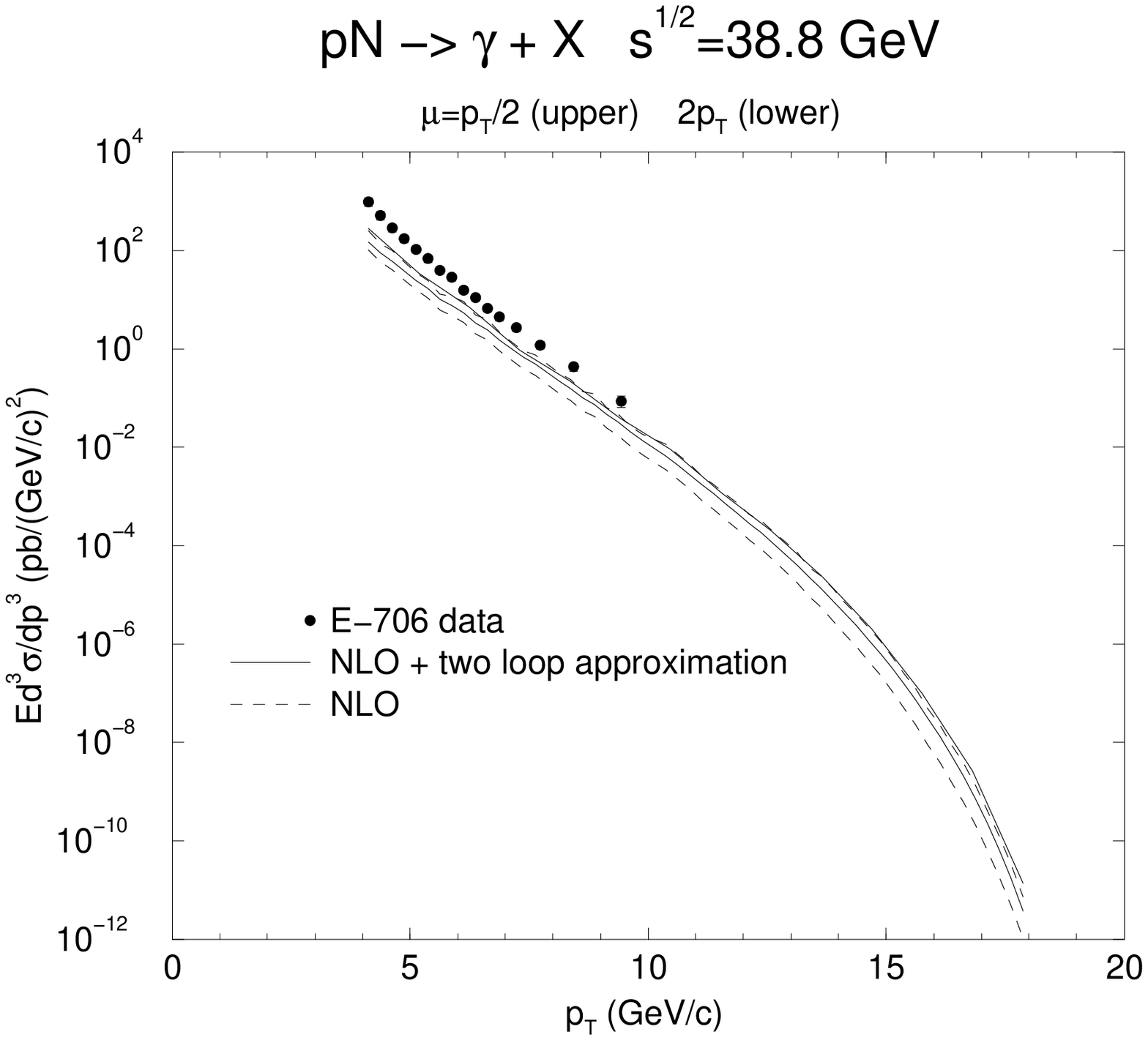}\\
\end{tabular}
\caption{Resummed calculations of isolated photon production compared
with the E706 data; 
left, by Catani {\it et al.}\protect\cite{cataniresum706} and 
right, by Kidonakis\protect\cite{kidoresum706}.
\label{fig:resum}}
\end{center}
\end{figure}

Much larger deviations from QCD are observed in fixed-target experiments
such as E706 at Fermilab\cite{e706}.  Again, Gaussian smearing (with 
$k_T \approx 1.2$~GeV in this case) can account for the data, as shown
in Fig.~\ref{fig:kt706}.
A rather different view is expressed by Aurenche and 
collaborators\cite{aurenche},
who find their calculations, {\it sans} $k_T$, to be consistent with all data 
with the sole exception of E706 (see Fig.~\ref{fig:aurenche}).  
They say ``it does not appear very 
instructive to hide this problem by introducing an arbitrary parameter
fitted to the data at each energy.''  
Indeed, I believe that Gaussian smearing has told us pretty much all that 
it can.  Its predictivce power is small: what happens to forward photons,
for example?  The ``right way'' to treat soft gluon emission should be 
through a resummation calculation which works nicely for $\gamma\gamma$ and
$W/Z$ transverse momentum distributions.  Unfortunately, this approach
does not seem to model the E706 data, which still lie a factor of two
or more above the resummed QCD 
calculations\cite{cataniresum706}\cite{kidoresum706} (Fig.~\ref{fig:resum}).  
I do not know if there are
other terms that can be resummed, of whether this should be taken as
an implication that the whole idea of soft gluons being responsible 
for this discrepancy is mistaken.

\begin{figure}[tb]
\begin{center}
\includegraphics*[height=6.5cm]{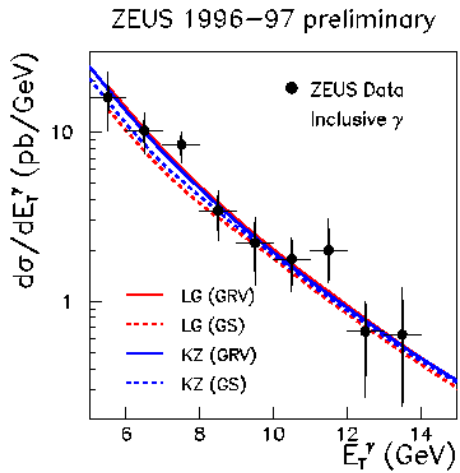}
\caption{Isolated photon cross section as measured by ZEUS, compared
to the NLO QCD prediction\protect\cite{zeusphotons}.
\label{fig:zeusgamma}}
\end{center}
\end{figure}

As an interesting aside, we may note that the ZEUS measurement
of prompt photon production at HERA\cite{zeusphotons}, 
which covers a similar range
of transverse momenta to E706, agrees well with NLO QCD without any need 
for $k_T$ (Fig.~\ref{fig:zeusgamma}).  
It is true that the agreement is only good if the ``resolved'' 
photon contribution to the DIS process is included, and this does
form a kind of resummation.  It has also been pointed out that the 
enhancements from  $k_T$ would be smaller here than for E706 as
the cross section is less steeply falling, with only a 20\% enhancement
expected at $E_T = 5$~GeV.

In summary, direct photon production has proved extremely interesting
and remains quite controversial. The appropriateness of a Gaussian $k_T$
treatment is still hotly debated, the experiments may not all be 
consistent, and resummation has not proved to be the answer.  

\section{Weak Bosons}
\index{Z boson}
\index{W boson}

\subsection{$Z$ Transverse Momentum}

\begin{figure}[tb]
\begin{center}
\includegraphics*[height=6.5cm]{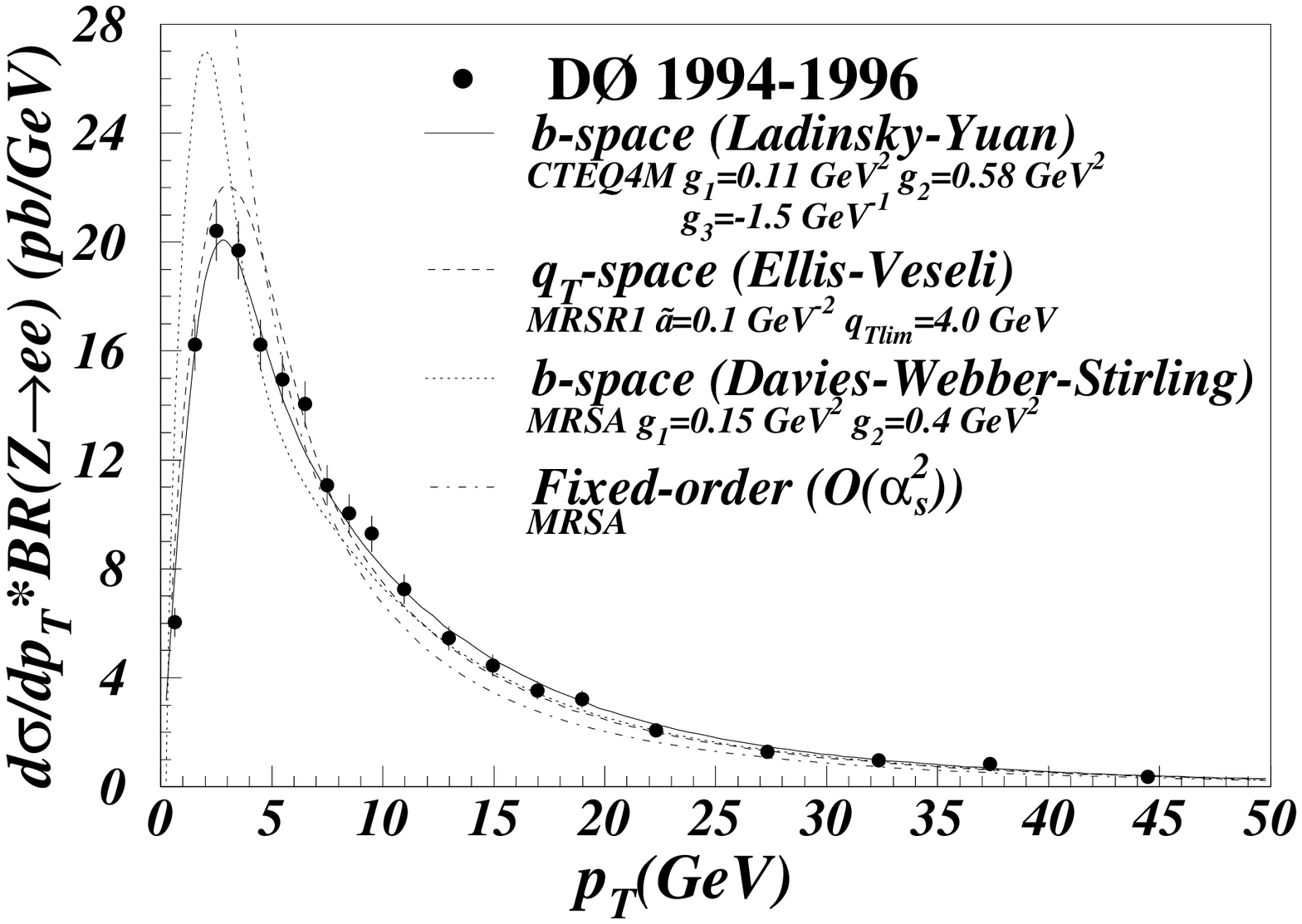}
\begin{tabular}{cc}
\includegraphics*[height=4cm]{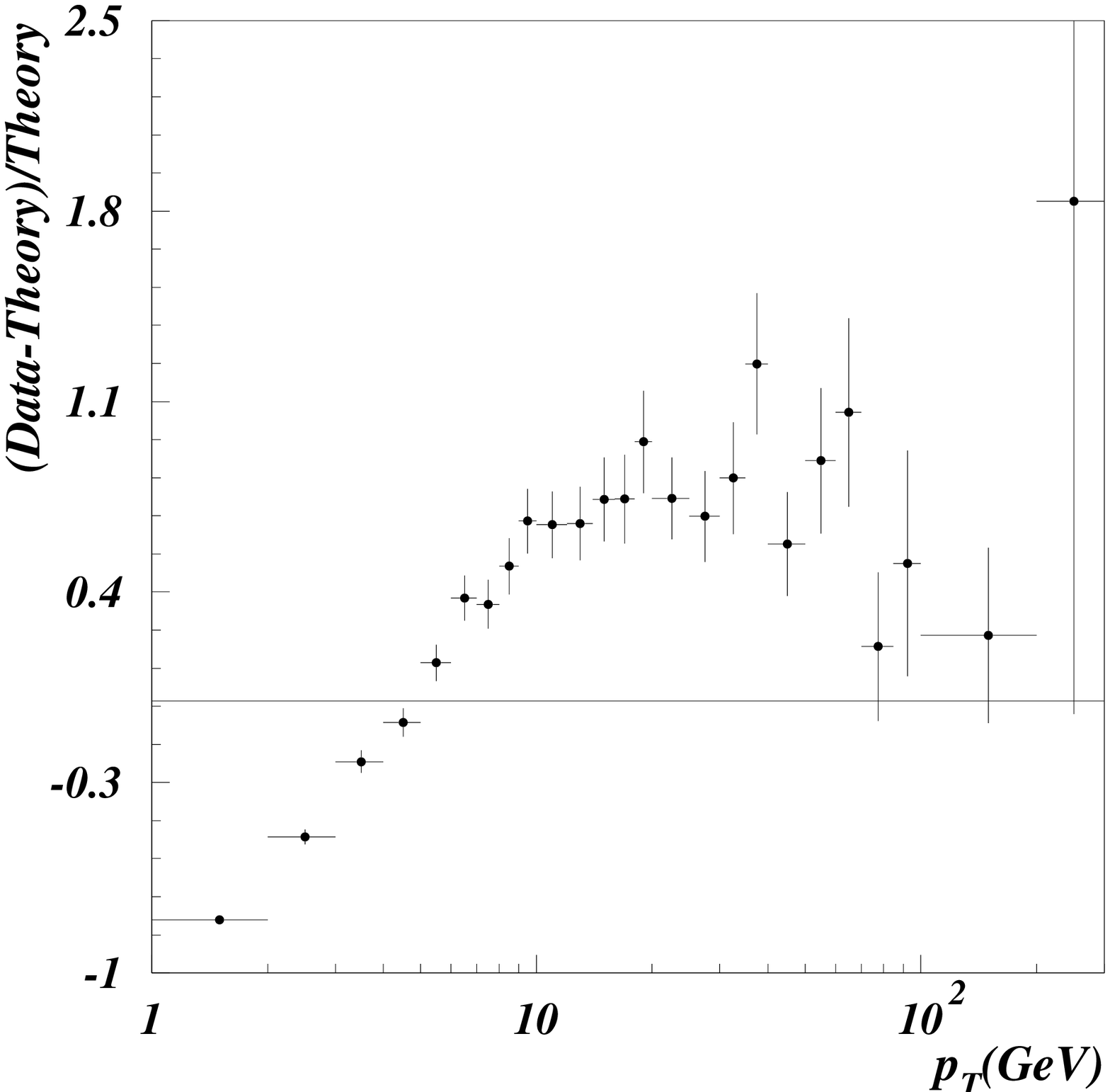} &
\includegraphics*[height=4cm]{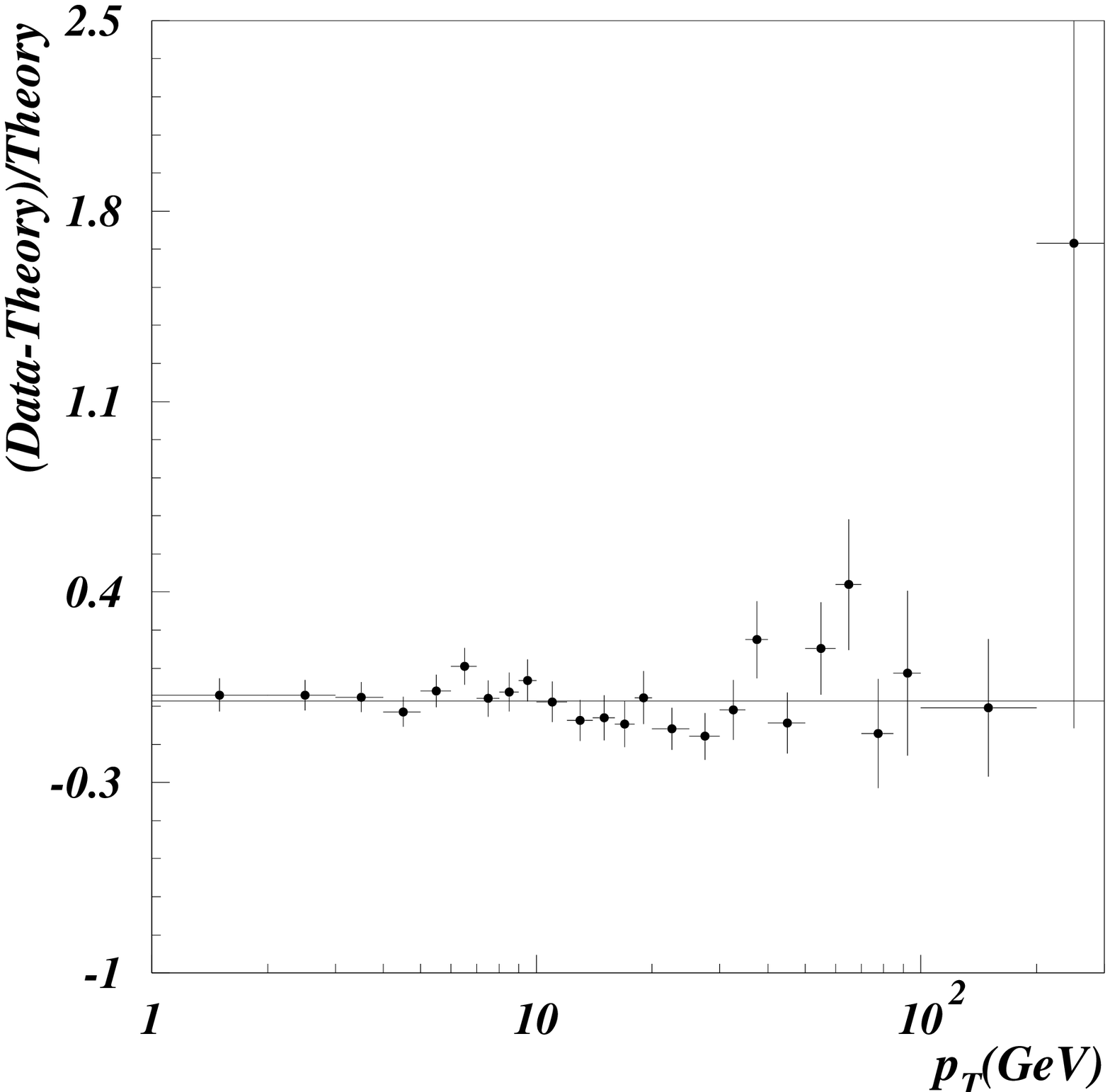} \\
\end{tabular}
\caption{Transverse momentum distribution of the $Z$, as measured by
D\O\protect\cite{d0ptz}. 
The upper plot shows the data and various calculations.  The
lower left shows the data normalized to the fixed-order QCD prediction and
the lower right shows the data normalized to the resummed calculation
of Ladinsky and Yuan\protect\cite{ladinsky}.    
\label{fig:ptz}}
\end{center}
\end{figure}

D\O\ has new results on the transverse momentum distribution
of the $Z$ boson\cite{d0ptz}.  Figure~\ref{fig:ptz} shows the
data compared with a variety of QCD predictions.  Clearly the
fixed-order NLO QCD is not a good match for the data, while the
resummed formalism of Ladinsky and Yuan\cite{ladinsky} 
fits rather well.  On the
other hand the resummed calculations of 
Davies, Webber and Stirling\cite{dws}, 
and of
Ellis and Veseli\cite{eandv}, 
do not offer quite as good a description of the data.

\subsection{$W+$jets}

D\O\ used to show a cross section ratio $(W+1{\rm jet})/(W+0{\rm jet})$
which was badly in disagreement with QCD.  This is no longer shown:
the data were basically correct,  but there was a bug in the way
D\O\ extracted rhe ratio from the DYRAD theory calculation.

\begin{figure}[tb]
\begin{center}
\begin{tabular}{cc}
\includegraphics*[height=6.5cm]{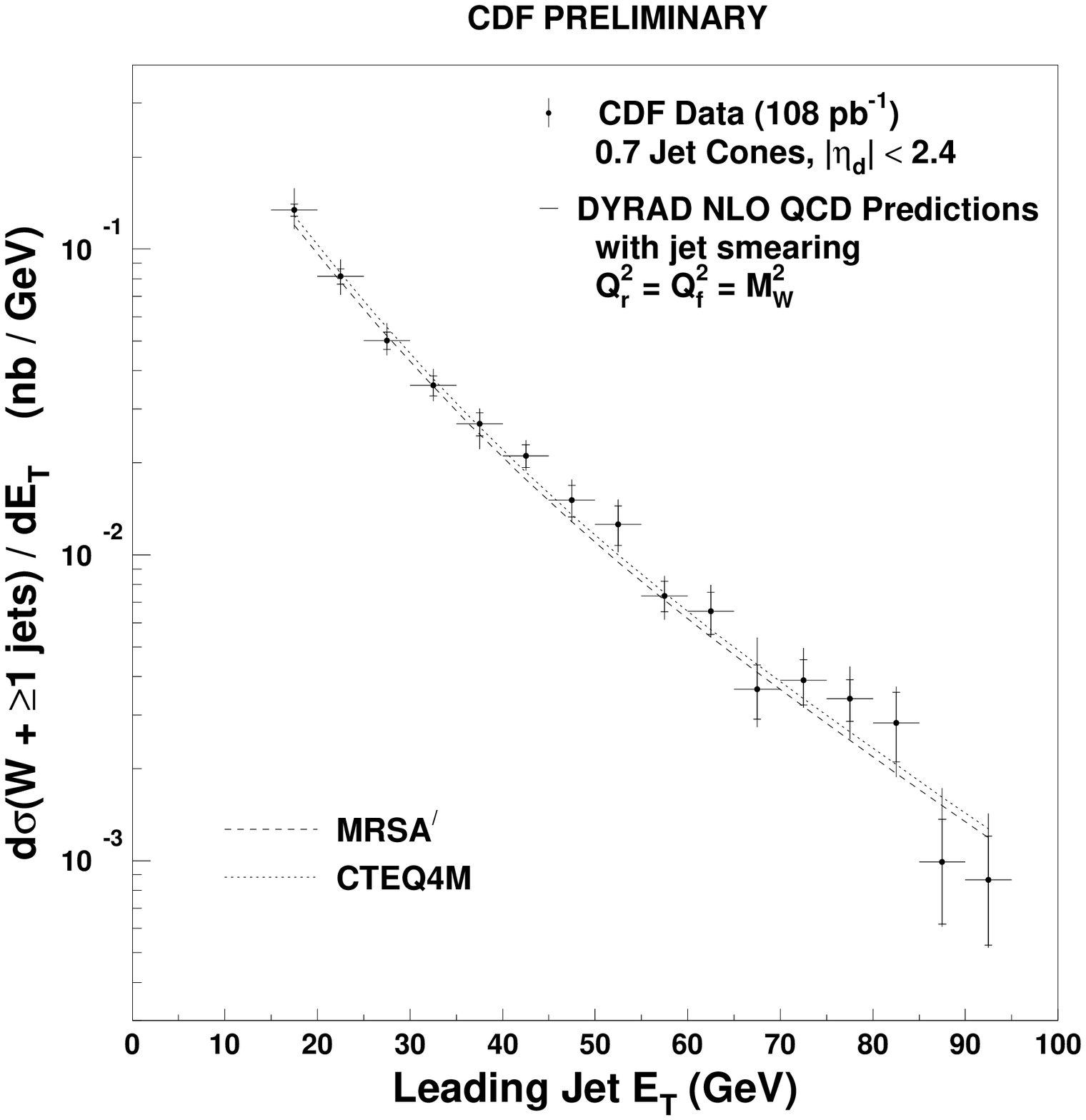}&
\includegraphics*[height=6.5cm]{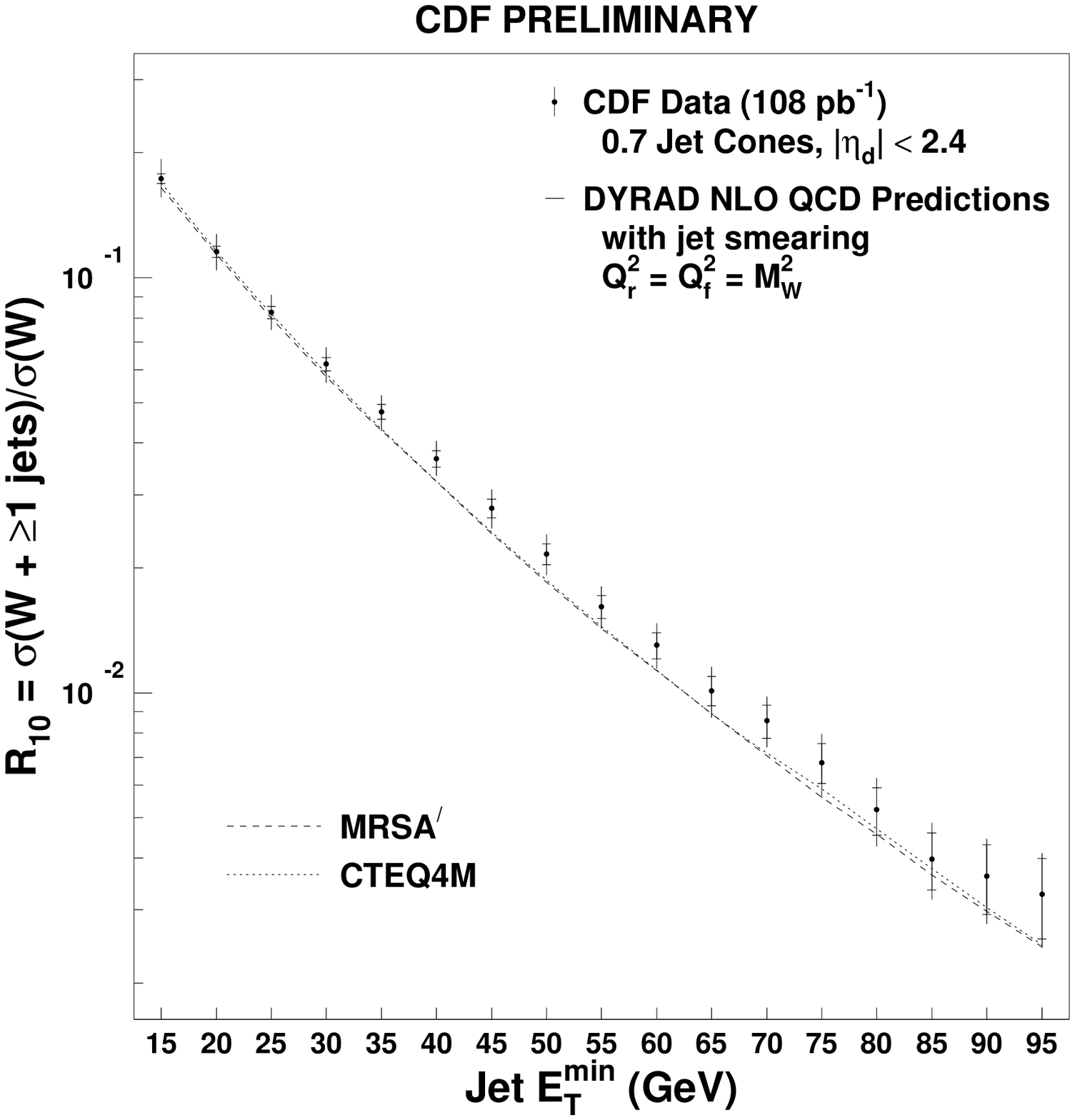}\\
\end{tabular}
\caption{The left hand plot shows the inclusive cross section for
$W+$jets production at the Tevatron, as measured by 
CDF\protect\cite{cdfwjets}.  The right
hand plot shows the ratio of the $W+$jets cross section to the inclusive
$W$ cross section.
\label{fig:cdfwjets}}
\end{center}
\end{figure}

Recent CDF measurements of the $W+$jets cross sections\cite{cdfwjets} 
agree well with
QCD, as shown in Fig.~\ref{fig:cdfwjets}.

\section{Heavy Flavour Production}
\index{Heavy flavour production}

\begin{figure}[tb]
\begin{center}
\begin{tabular}{cc}
\includegraphics*[height=6.5cm]{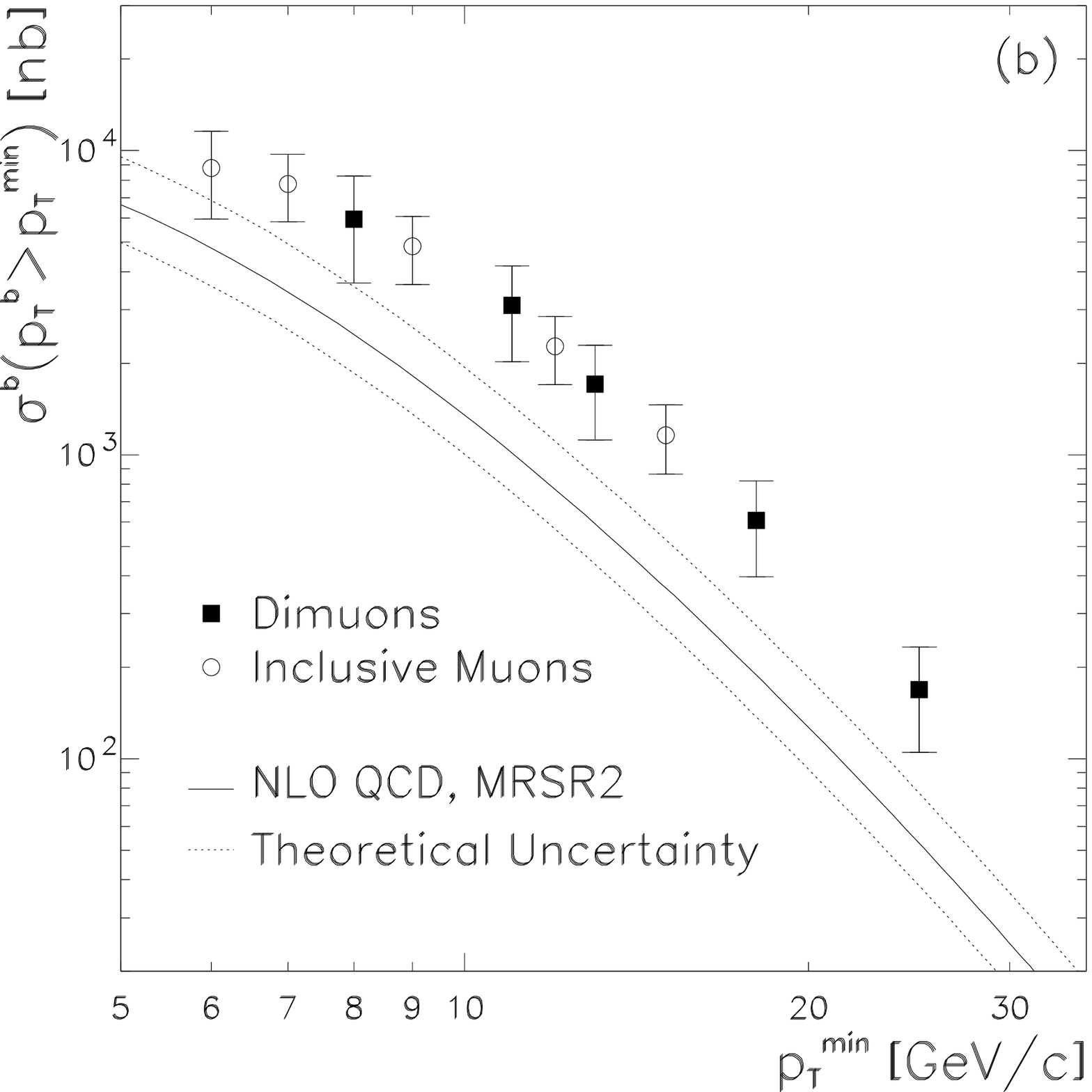}&
\includegraphics*[height=6.5cm]{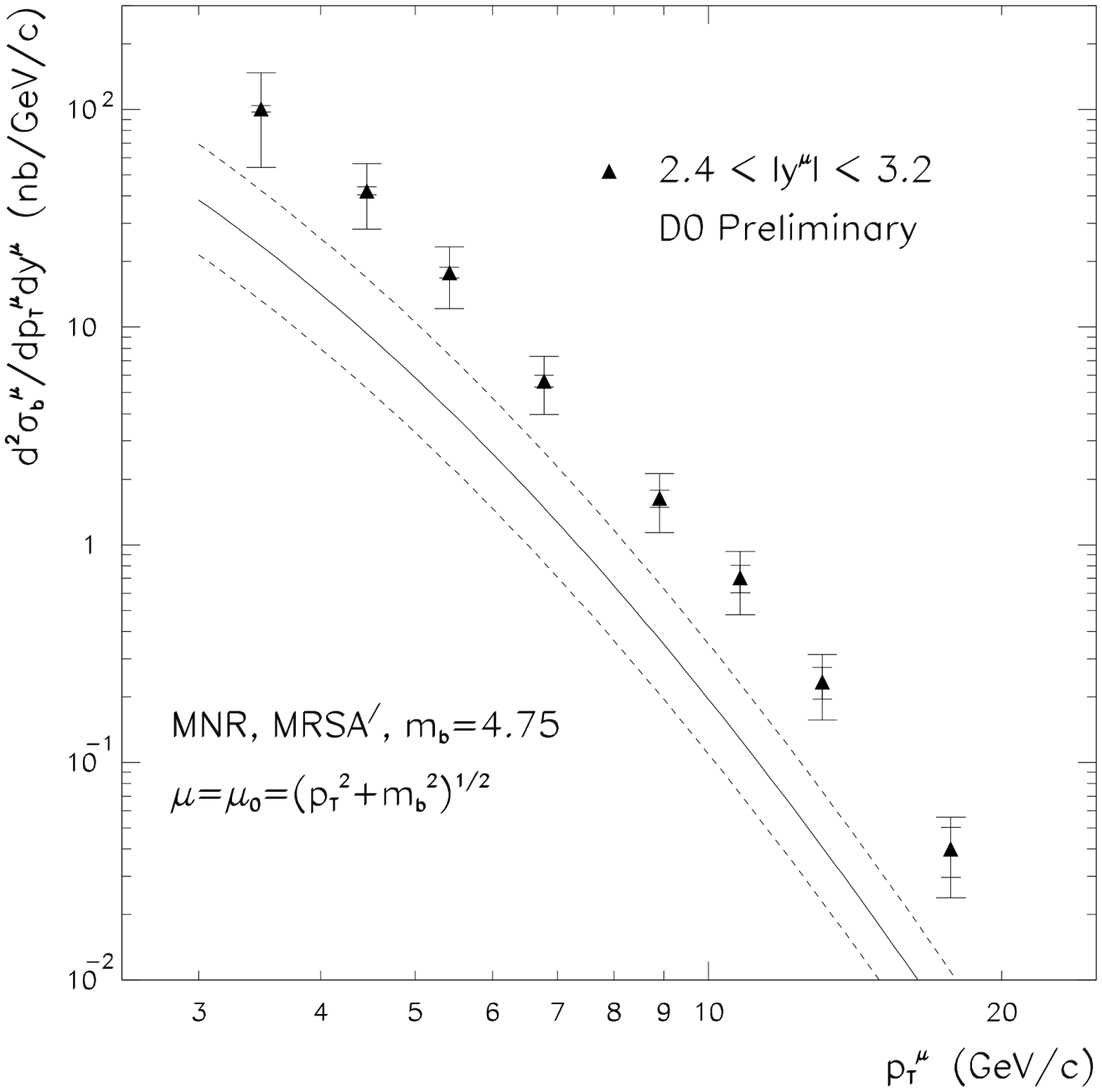}\\
\end{tabular}
\caption{Cross sections for $b$ production at the Tevatron compared
with NLO QCD predictions, as measured by D\O\protect\cite{d0b};
left, central rapidity region, and
right, forward.
\label{fig:d0b}}
\end{center}
\end{figure}

At the Tevatron, the measured inclusive $b$ and $B-$meson production cross
sections continue to lie a factor of about two above the NLO QCD 
expectation.  This is seen by both CDF\cite{cdfb} and D\O\cite{d0b} 
in the central and
forward regions (the difference is perhaps even larger for
forward $b$ production, as seen in Fig.~\ref{fig:d0b}).  
On the other hand, QCD does a good job
of predicting the shape of inclusive distributions, and of the
correlations between $b$ quark pairs, so it seems unlikely that
any exotic new production mechanism is responsible for the higher
than expected cross section.

\begin{figure}[tb]
\begin{center}
\begin{tabular}{cc}
\includegraphics*[height=6.5cm]{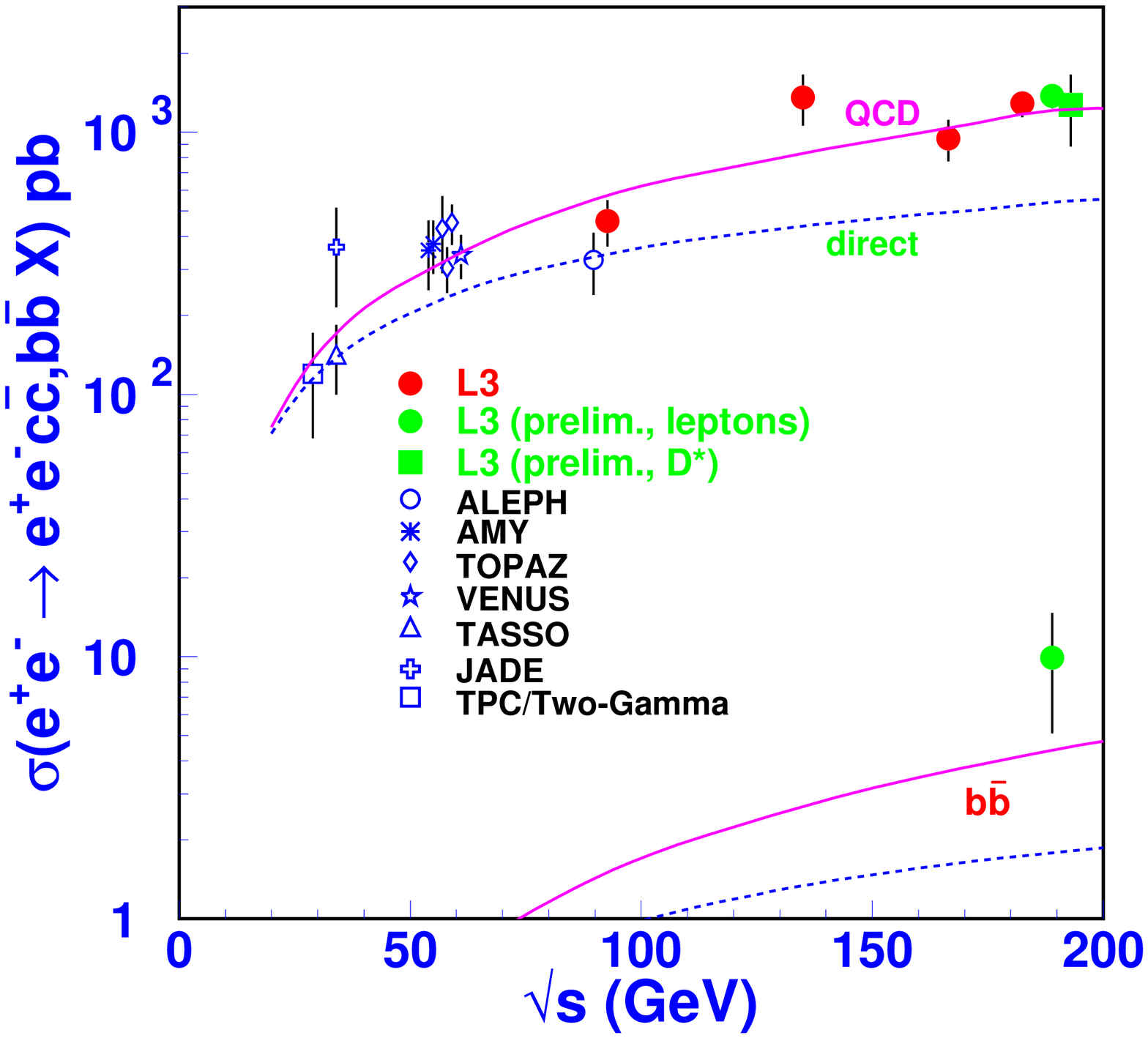}&
\includegraphics*[height=4cm]{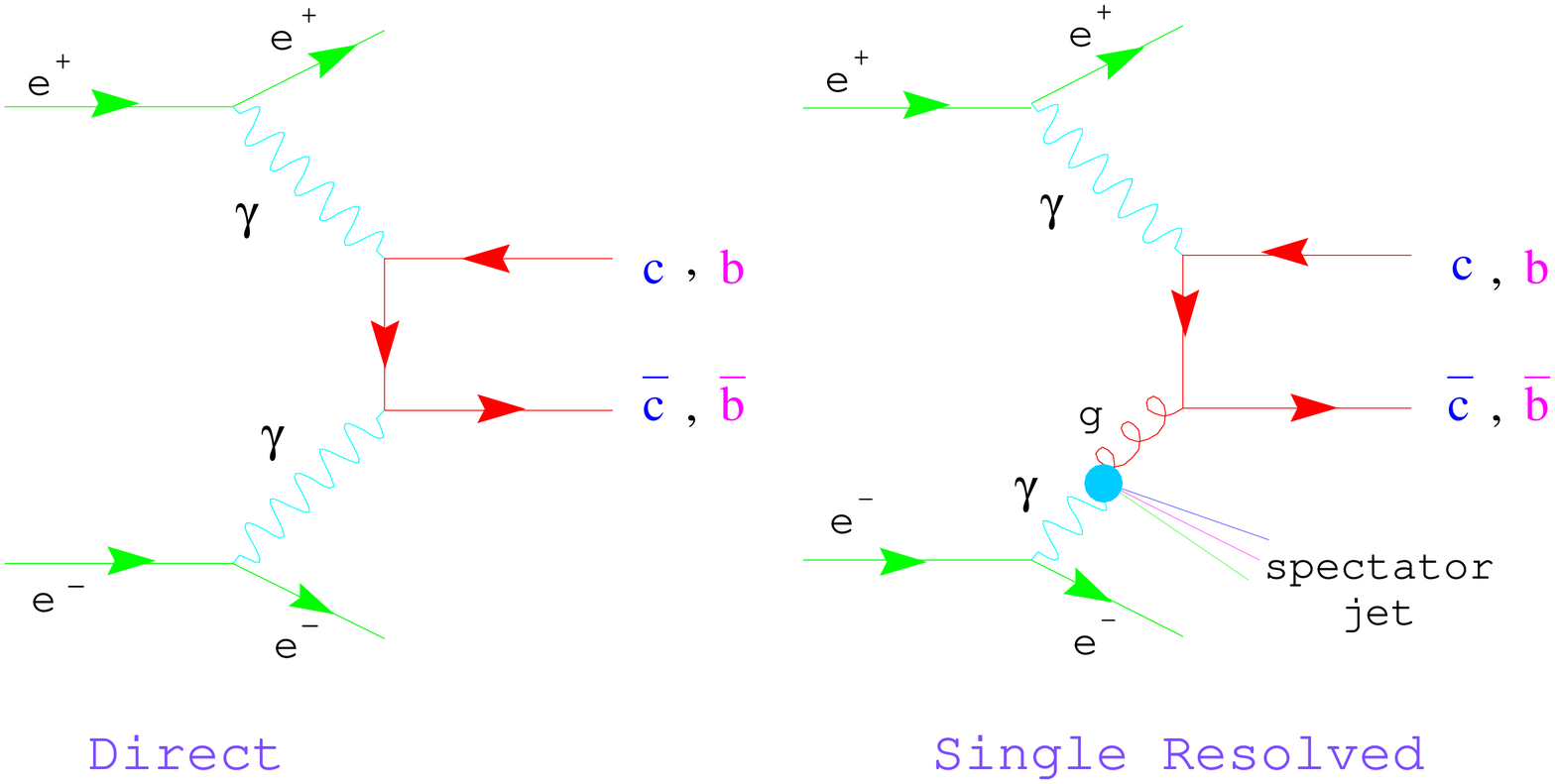}\\
\end{tabular}
\caption{Cross sections for heavy quark production measured at 
LEP\protect\cite{l3bquark}.
The Feynman diagrams show the direct and resolved photon contributions
to the process.
\label{fig:l3b}}
\end{center}
\end{figure}

\begin{figure}[tb]
\begin{center}
\begin{tabular}{cc}
\includegraphics*[height=7.0cm]{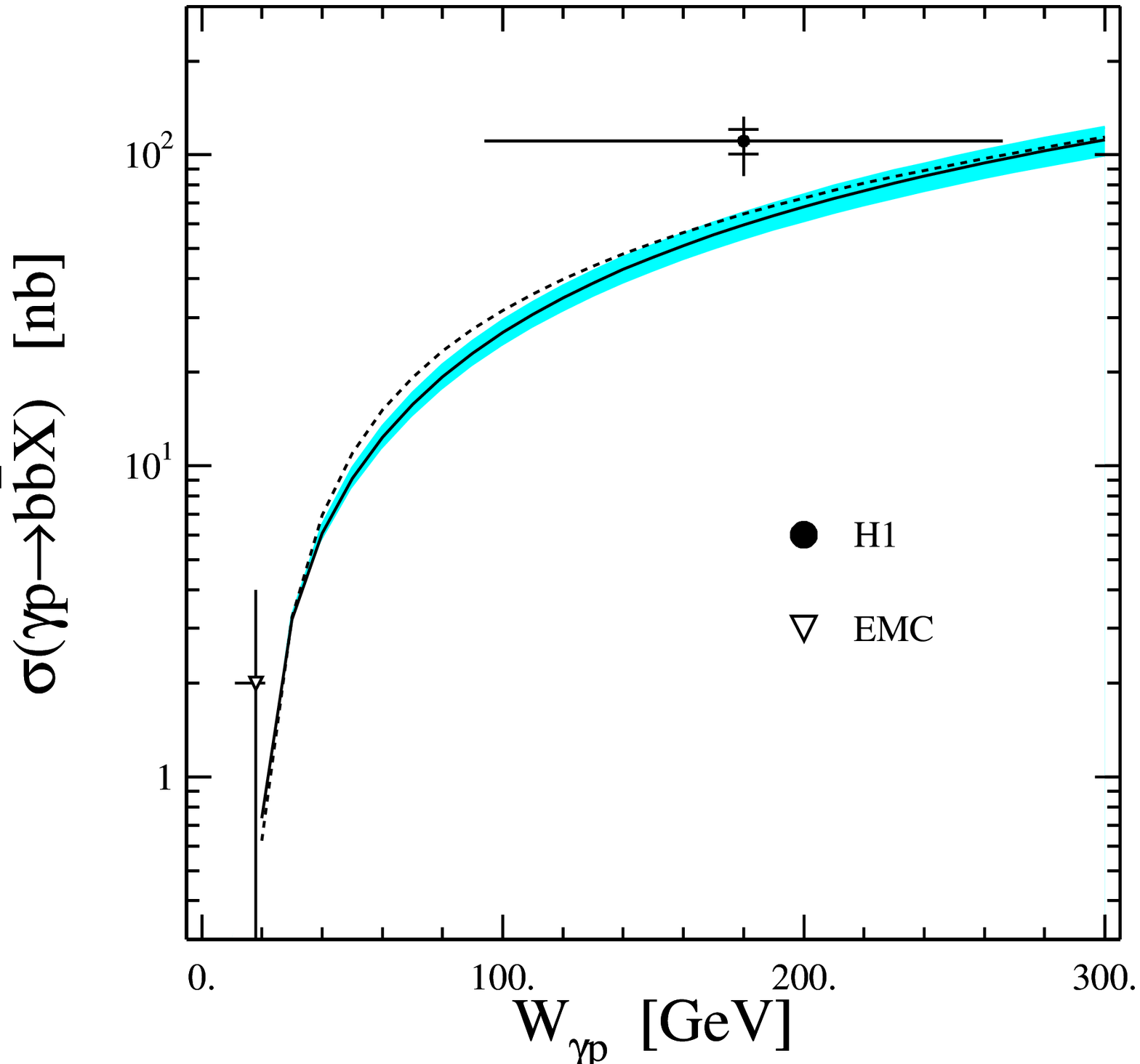}&
\includegraphics*[height=6.5cm]{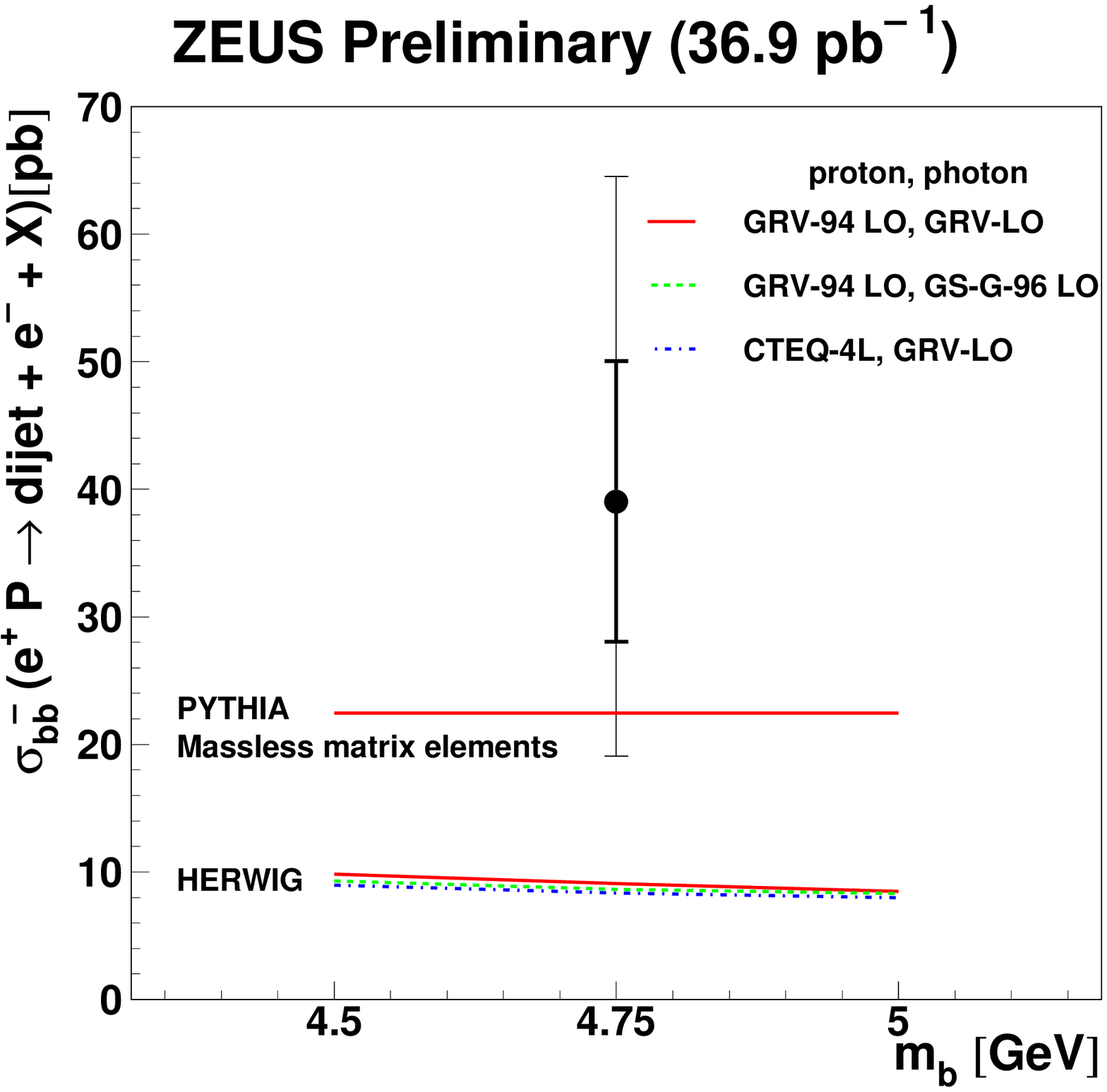}\\
\end{tabular}
\caption{H1\protect\cite{h1bquark} 
and ZEUS\protect\cite{zeusbquark} measurements of 
the $b$ production cross section at HERA,
compared to NLO and LO QCD respectively.
\label{fig:zeusb}}
\end{center}
\end{figure}

There are now results on heavy quark production at LEP 2.  In
particular, L3 has reported~\cite{l3bquark} the first observation
of $b$ production in $\gamma\gamma$ collisions 
($e^+e^- \to e^+e^-b\overline b$).  As shown in Fig.~\ref{fig:l3b}
the cross section is, once again, 2--3 times the QCD expectation.
The picture is rather similar at HERA.  H1\cite{h1bquark} and 
ZEUS\cite{zeusbquark} have reported
$b$ production cross sections in $\gamma p$ collisions, 
using leptons to tag $b$-jets.  As shown in Fig.~\ref{fig:zeusb}
the cross section is at least a factor of two above the QCD
prediction.
We therefore have a picture of consistent experimental results which
are unfortunately all inconsistent with theory!  

If the heavy flavour is heavy enough, QCD seems to work rather better.
The current state of measured and predicted top cross sections is
summarised in Table~\ref{tab:top}.  This includes the latest (revised downward)
CDF measurement.  There is an excellent agreement between data and
theory.

\index{top}

\begin{table}[tb]
\begin{center}
\begin{tabular}{|l|c|}  
\hline
Authors & Cross Section (pb)\\
\hline
\hline
CDF\protect\cite{cdftop}  & $6.5^{+1.7}_{-1.4}$ (at $m_t=175$~GeV)\\
D\O\protect\cite{d0top}   & $5.9\pm 1.7$ (at $m_t=172$~GeV)\\
\hline  
Bonciani {\it et al.}\protect\cite{boncianitop} & $5.0\pm 1.6$\\
Berger and Contopanagos\protect\cite{bergertop} & $5.6^{+0.1}_{-0.4}$\\
Kidonakis\protect\cite{kidonakistop} & $7.0$\\
\hline
\end{tabular}
\caption{Top production cross sections at the Tevatron, measured and
predicted.}
\label{tab:top}
\end{center}
\end{table}

\begin{figure}[p]
\begin{center}
\epsfig{file=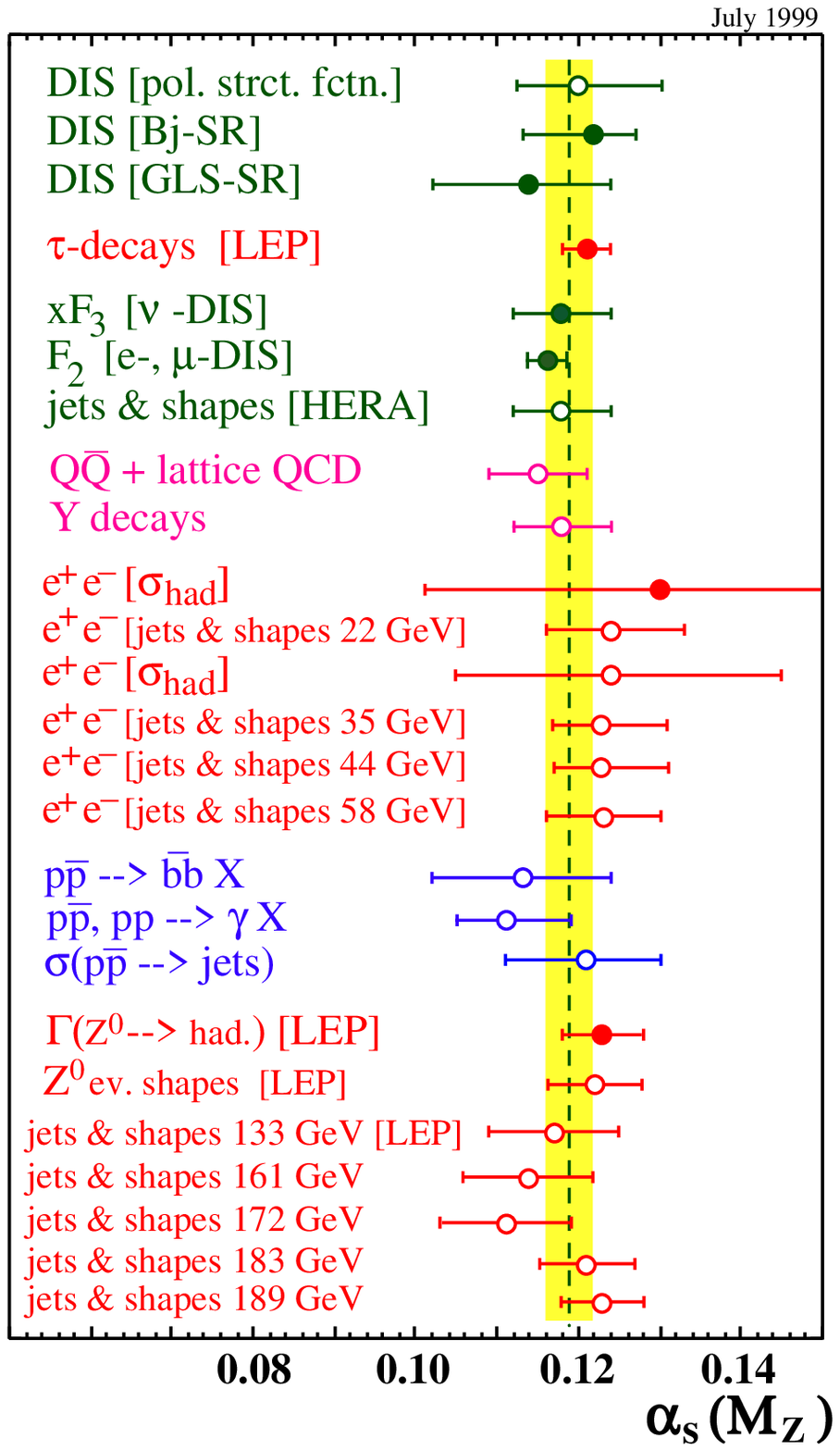}
\caption{Compilation of $\alpha_s(m_Z)$ measurements. The summer 1999
world average value is $0.119 \pm 0.003$
\protect\cite{bethkeprivate}.}
\label{fig:bethke}
\end{center}
\end{figure}

\section{Measurements of $\alpha_s$}
\index{$\alpha_s$}

The strong coupling $\alpha_s$ is a fundamental parameter of QCD.
Its value cannot be calculated, but must be determined experimentally.
A number of new measurements have been reported recently.

Deep inelastic scattering data is now being interpreted in an 
NNLO framework.  Santiago and Yndur\'ain report\cite{santiago}
an extraction of $\alpha_s$ from $F_2$ measured at SLAC, BCDMS,
E665 and HERA; they obtain
$\alpha_s(m_Z) = 0.1163 \pm 0.0023$.
Kataev, Parente and Sidorov\cite{kataev} extracted 
$\alpha_s$ from $xF_3$ measured at CCFR and obtain
$\alpha_s(m_Z) = 0.118 \pm 0.006$.

The LEP electroweak working group has reported\cite{quast} new values
from LEP~1/SLD $Z$ pole data.  The full Standard Model fit
yields a value of
$\alpha_s(m_Z) = 0.119 \pm 0.003$
while the ratio of the $Z$ partial widths to hadrons and leptons gives
$\alpha_s(m_Z) = 0.119 \pm 0.004 ^{+0.003}_{-0}(m_H)$.
The LEP experiments have all extracted $\alpha_s$ from event shapes,
charged particle and jet multiplicities
at $\sqrt{s}=130-196$~GeV.  Monte Carlo programs are
used to model non-perturbative effects. 
Both L3\cite{l3alphas} and OPAL\cite{opalalphas} 
have nicely demonstrated the running of $\alpha_s$; in the case
of L3, by using radiative events to access a lower $\sqrt{s}$, and
in OPAL's case by combining their data with that of JADE.
Typical uncertainties on $\alpha_s(m_Z)$ from the LEP2 data are
around $\pm 0.006$.

At HERA, 
$\alpha_s$ has been extracted 
from the inclusive jet rate and the dijet rate (H1\cite{h1alphas})
and the dijet fraction (ZEUS\cite{zeusalphas}).  The values 
of  $\alpha_s(m_Z)$ obtained
are consistent with the world average, with uncertainties of
around $\pm 0.005-0.008$.

S.~Bethke\cite{bethkeprivate} 
has kindly provided me with an updated world average
value of $\alpha_s(m_Z)$ for this meeting.  It is based
on 25 measurements listed in Fig.~\ref{fig:bethke}:
$$\alpha_s(m_Z) = 0.119 \pm 0.004.$$
If one uses only complete NNLO QCD results (the filled symbols
in Fig.\ref{fig:bethke}) one obtains:
$$\alpha_s(m_Z) = 0.119 \pm 0.003.$$
We note excellent consistency bwteeen low and high energy data,
and between deep inelastic scattering, electron-positron and
hadron collider data.  Changes from the previous world 
average\cite{bethke98} are minimal.

\begin{figure}[tb]
\begin{center}
\includegraphics*[height=8cm]{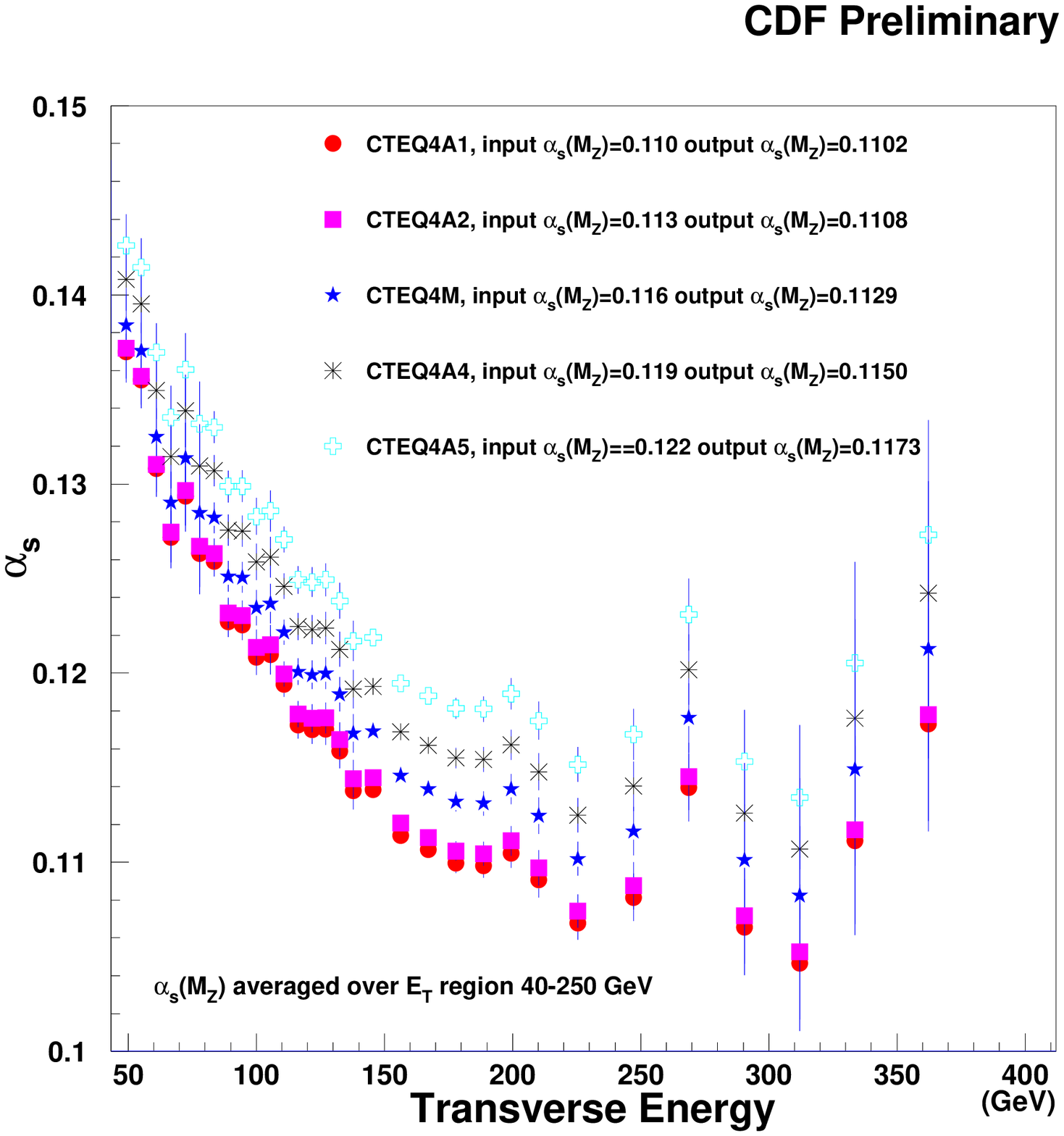}
\caption{Value of $\alpha_s$ as a function of scale (jet transverse 
energy) inferred by CDF from the jet cross section using the CTEQ4A
series of parton distributions\protect\cite{cdfalphas}.}
\label{fig:cdfalphas}
\end{center}
\end{figure}

\subsection{Consistency Tests}

At this point we know the value of $\alpha_s$ rather well --- it is hard to 
point to a prediction which is limited by its precision.  Hence some of
the more interesting measurements are really tests of self-consistency 
and of our understanding of QCD, 
rather than determinations of this parameter. 

An example is the use of power corrections for the
non-perturbative corrections to event shape variables, as described in
Bryan Webber's contribution to these proceedings.  I would place
DELPHI's extraction of $\alpha_s$ 
from oriented event shape variables\cite{oes} in the same category.
Fits at $m_Z$ yield very precise values of $\alpha_s$: $0.1180\pm 0.0018$
from the jet cone energy fraction, for example.  However, the fitting
procedure relies on ``optimization'' of the renormalization scale for each
variable through a simultaneous fit of $\alpha_s$ and $x_\mu = \mu^2/Q^2$.
For 18 jet shape variables the resulting scales range from 
$\mu^2 = (0.003Q)^2$ to $(7Q)^2$, a much larger range than one is
comfortable with.  In fact the whole procedure has been called 
theoretically unjustified.  Nonetheless the consistency of the results
is certainly interesting:  if $x_\mu$ is fixed to 1, the spread in
the extracted $\alpha_s$ values is much larger.  I suspect that this is
telling us something about QCD (though maybe not, or not just,
the value of $\alpha_s$). 

Another measurement of this type is the extraction of $\alpha_s$ from 
the CDF inclusive jet cross section\cite{cdfalphas}.
The value quoted, 
$\alpha_s(m_Z) = 0.113^{0.008}_{-0.009}$, is consistent with the world
average, and $\alpha_s$ shows a nice evolution with scale (given by the
jet transverse energy), as shown in Fig.~\ref{fig:cdfalphas}.  
However the figure also shows that the measurement suffers from a large, 
and hard to quantify, sensitivity on the parton distributions,
especially on the value of $\alpha_s$ assumed therein.  
At this time I think it must be characterized as a nice test of QCD and not
a measurement of $\alpha_s$.

\section{Some Final Remarks}

In closing I will take the opportunity to outline some areas where
I think progress would be welcome: ``What I would like for Christmas''.

Firstly parton distributions with quoted uncertainties, or at least with
a technique
for the propagation of uncertainties as outlined by Giele and collaborators.
This would spare us from future unnecessary excitement over things like
the high-$E_T$ jet ``excess.''

Secondly I would like to see a theoretical and experimental effort to 
understand the underlying event in hadronic collisions.  It is an 
inconsistent treatment of the event to subtract it out of the jet
energies, as is usually done.  The ratio of 1800 to 630~GeV jet cross
sections may indicate problems with this approach.  And such
understanding would also enable a consistent treatment of double
parton scattering which may be very important at the LHC.

\begin{figure}[tb]
\begin{center}
\includegraphics*[height=8cm]{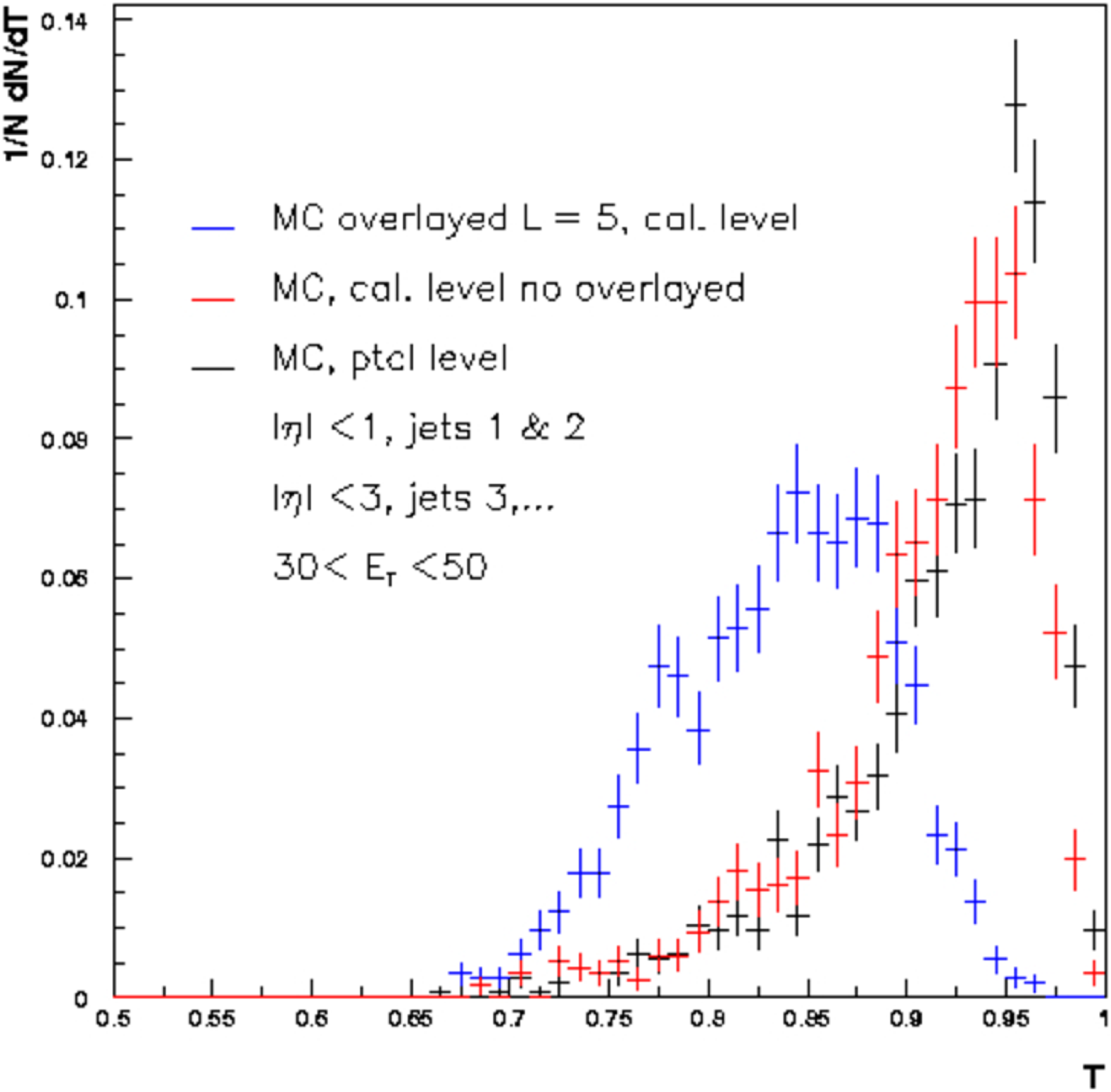}
\caption{The importance of noise to jet algorithms.  The figure shows
a D\O\ simulation of thrust distributions evaluated using $k_T$ jets.
The calorimeter measures a distribution close to the ideal, until
noise is added to the simulation, at which point it is dramatically
shifted to lower values. 
\label{fig:noise}}
\end{center}
\end{figure}

\index{jet algorithm}
Thirdly we need progress on jet algorithms for hadron colliders.
Indeed, there is a lot of work going on in workshops at Fermilab
and Les Houches.  There are various species of $k_T$ algorithm
to be compared, and the question remains whether the cone algorithm
can be made theoretically acceptable.  Theoertical requirements 
centre on the need for infrared and collinear safety, and the avoidance
of ad hoc parameters; experimentalists worry about sensitivity to noise,
pileup and negative energies. These can have potentially large
effects on jet measurements especially at low energies, as shown in 
Fig.~\ref{fig:noise}.

\index{hard diffraction}

\begin{figure}[tb]
\begin{center}
\begin{tabular}{cc}
\includegraphics*[height=6cm]{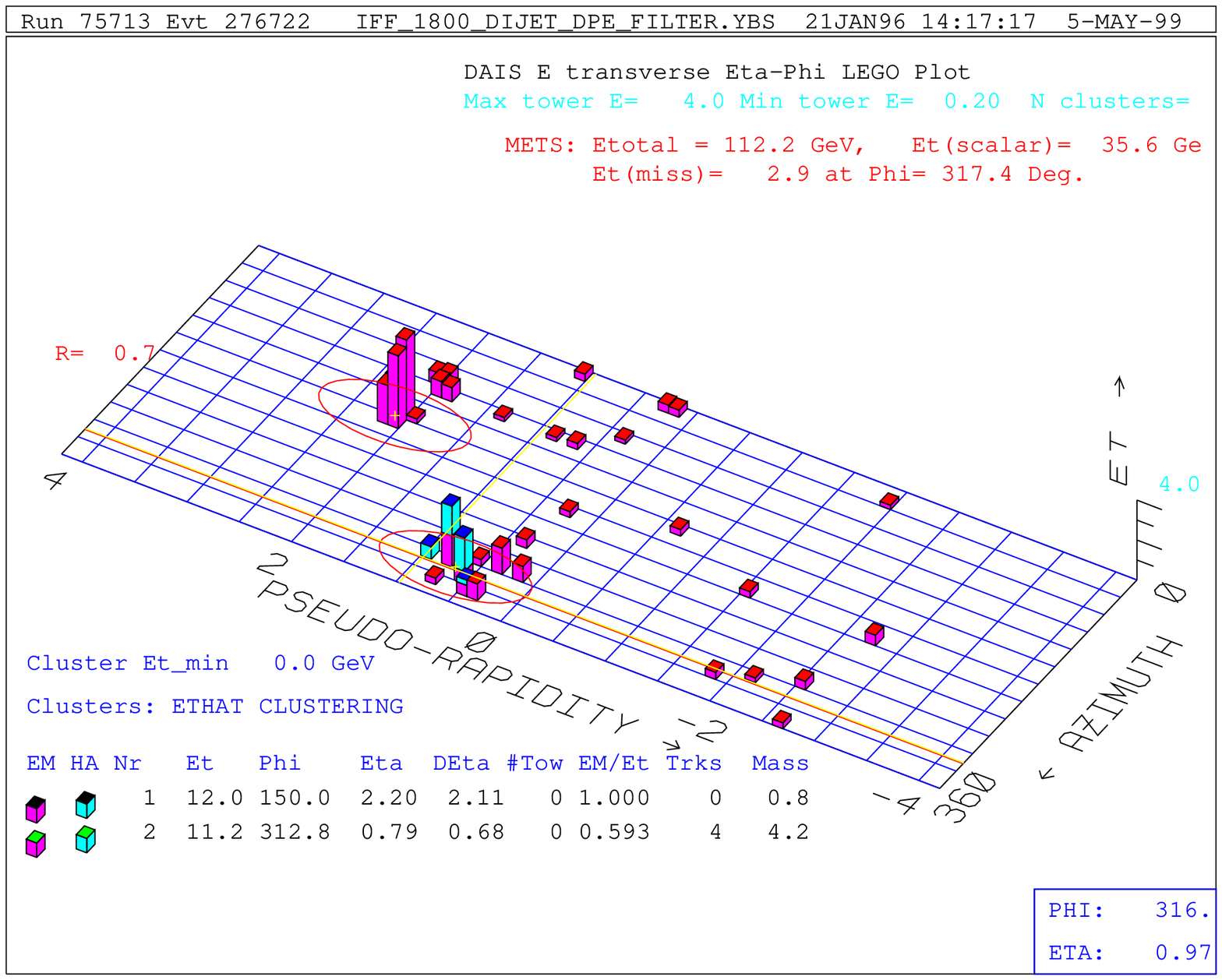} &
\includegraphics*[height=6cm]{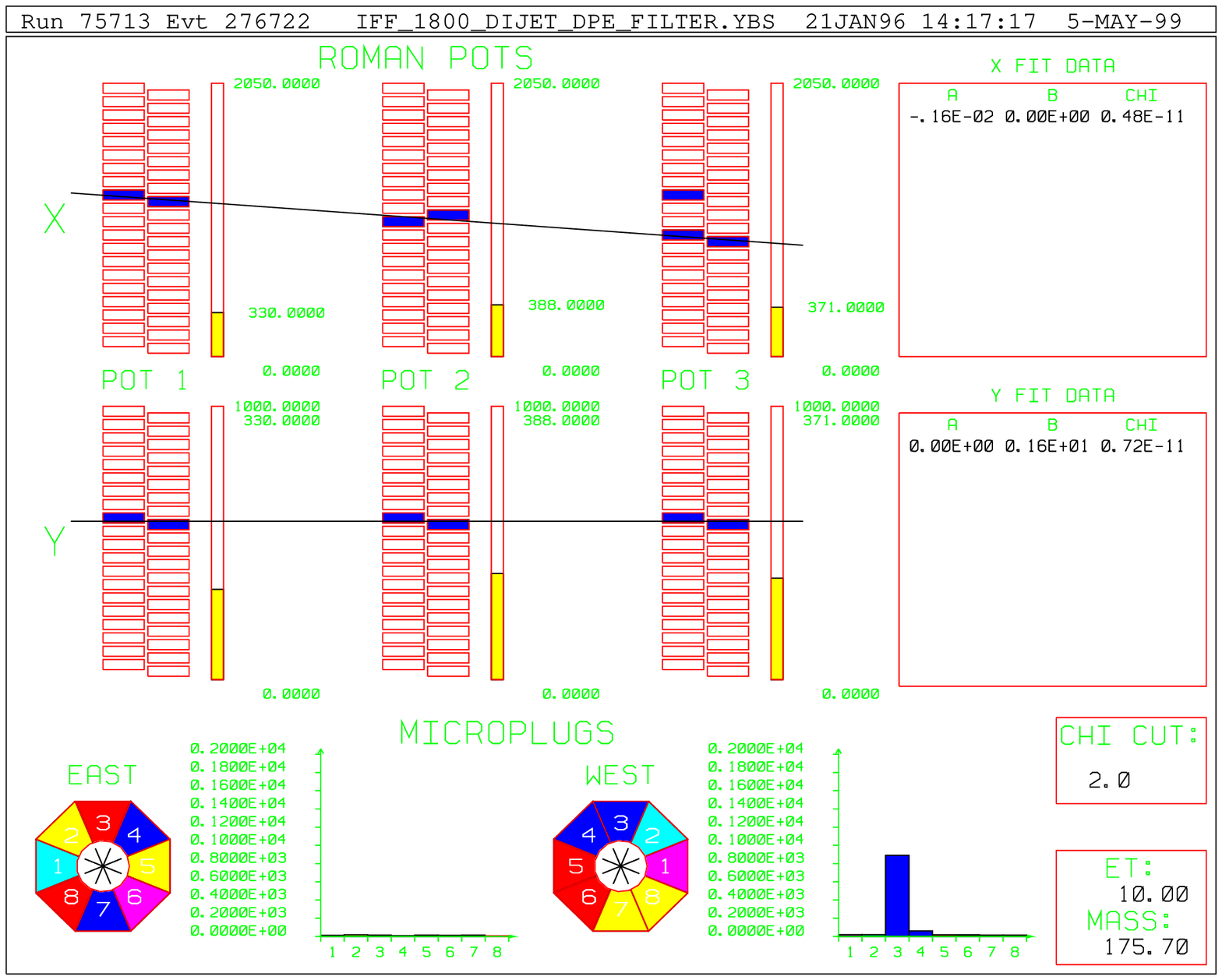} \\
\end{tabular}
\caption{CDF double pomeron exchange dijet candidate.
The left hand plot shows jets observed in the calorimeter; the right hand
plot shows the two tracks in the roman pot detectors which are
attributed to the undissociated proton and antiproton.
\label{fig:potevent}}
\end{center}
\end{figure}

Fourthly there is one QCD process that we have completely failed to
describe so far.  Figure~\ref{fig:potevent} shows a CDF event with
a track in the Roman Pot detectors.  It is jet production, a
perturbative process which I have claimed is well-modelled by
NLO QCD.  Except for one detail: in a substantial fraction (a few
percent?) of such events, one of the protons doesn't break up.
Whether we call this pomeron exchange, or think of it in terms of
parton correlations inside the proton, it doesn't form part of a
consistent description of hard hadronic interactions.  My hunch
is that we won't get too far in understanding processes like this
as long as we think of them as being somehow alien to perturbative
QCD.

\section{Conclusions}

In conclusion, testing QCD typically means testing our ability to
calculate within QCD.  In fact our calculational tools are working
quite well, especially at moderate to high energy scales.  The state
of the art is NNLO calculations, and NLL resummations.  However,
interesting things (challenges!) start happening as we reduce the energy
scale to of order 5~GeV.  We have problems calculating $b$ quark
cross sections; there are problems with low-$p_T$ direct photon
production ($k_T$?); and perhaps indications of effects of a few GeV
in jet energies.  In addition, there are other challenges for
the future:  identification of appropriate jet algorithms, 
understanding of the underlying event in hadron-hadron collisions,
understanding parton distribution uncertainties, and obtaining a 
consistent picture of hard diffraction processes.  With two new
hadron collider facilities coming on line in the next decade, we
can be assured of a vibrant future for perturbative QCD and jet physics.

\def\Discussion{
\setlength{\parskip}{0.3cm}\setlength{\parindent}{0.0cm}
     \bigskip\bigskip      {\Large {\bf Discussion}} \bigskip}
\def\speaker#1{{\bf #1:}\ }

\newpage
\Discussion

\speaker{Alex Firestone (NSF)}
Regarding the isolated inclusive photon distributions, 
how plausible is it that at
least some of the problem is due to signal extraction above background?
Different 
experiments have different calorimeter resolution and different criteria
for photon isolation.  Could this be the source of the inconsistencies?

\speaker{Womersley}
Having carried out these analyses myself at the Tevatron, I certianly
understand that there
are large systematic errors at low $E_T$ because of the big jet backgrounds. 
There are
also problems in computing  the effects of the isolation requirements on
the theoretical predictions.  But the difficulty pointed
out by Aurenche {\it et al.} is that the E706 and the ISR data do not agree.
I do not really want to speculate on why that might be so.

\speaker{Jonathan Butterworth (University College London)}
It's true that the photon structure function is a way to resum partonic
contributions.  But the partons from the photon are collinear, 
so the $k_T$ effect is not included in this calculation. 

\speaker{Paul S\"oding (DESY)}
One of your wishes for Christmas has already become true; 
there is an analysis by
Michael Botje which provides PDF's with uncertainties.  
The paper is available.

\speaker{Michel Davier (LAL, Orsay)}
I think that the $\alpha_s$ average you quoted from Siggi Bethke 
has a very conservative
uncertainty (0.003).  The 2 most precise determinations 
(from $\tau$ decays and the $Z$ width)
have uncertainties already at this level and the jet rate at 
LEP yields results approaching this accuracy.  These determinations have 
different systematics.  So I would rather think that
the combined uncertainty is 0.002 at most.

\speaker{Siggi Bethke (Aachen and MPI Munich)}
Most of the individual errors quoted are lower limits, 
because there is some freedom to adjust the 
theoretical uncertainties.   The new world average that was quoted
relies, for the first time,  on NNLO calculations
alone, and also assumes some degree of correlation between
individual errors.  The scatter of the central values of the results is also
$\pm$ 0.003.

\end{document}